\documentclass[oneside]{elsarticle}

\usepackage{amsmath}
\usepackage{lineno}
\usepackage{setspace}
\usepackage{amssymb}
\usepackage{graphicx}
\makeatletter
\def\ps@pprintTitle{%
  \let\@oddhead\@empty
  \let\@evenhead\@empty
  \let\@oddfoot\@empty
  \let\@evenfoot\@oddfoot
}
\makeatother

\newcommand{\bx}{\mathbf{x}}

\newcommand{\bv}{\mathbf{v}}
\newcommand{\vc}{\mathbf{c}}
\newcommand{\ba}{\mathbf{a}}
\newcommand{\bn}{\mathbf{n}}
\usepackage[usenames]{color}

\newcommand{\pd}[2]{\frac{\partial #1}{\partial #2}}

\newcommand{\bc}{\mathbf{c}}
\definecolor{red}{rgb}{1,0,0.0}

\usepackage[margin=1.0in]{geometry}
\biboptions{numbers,sort&compress}

\usepackage{soul}
\usepackage{hyperref}
\RequirePackage{xcolor}					
\definecolor{linkcolor}{rgb}{0,0.4470,0.7410}

\hypersetup{
		bookmarks=true, bookmarksnumbered=true, bookmarksopen=true,
		unicode=true, colorlinks=true, linktoc=all, 
		linkcolor=linkcolor, citecolor=linkcolor, filecolor=linkcolor, urlcolor=linkcolor,
		pdfstartview=FitH, linkbordercolor=red,
}
\usepackage{color,soul}

\newcommand{\myhy}[2]{\href{#1}{\color{linkcolor}\setulcolor{linkcolor}\textbf{\ul{#2}}}}
\usepackage{etoolbox}
\patchcmd{\MaketitleBox}{\footnotesize\itshape\elsaddress\par\vskip36pt}{\itshape\elsaddress\par\parbox[b][36pt]{\linewidth}{\vfill\hfill\textnormal{Test}\hfill\null\vfill}}{}{}%
\patchcmd{\pprintMaketitle}{\footnotesize\itshape\elsaddress\par\vskip36pt}{\itshape\elsaddress\par\parbox[b][36pt]{\linewidth}{\vfill\hfill\textnormal{\textbf{Accepted manuscript version of} \myhy{https://doi.org/10.1016/j.jcp.2023.112472}{DOI:10.1016/j.jcp.2023.112472} }\hfill\null\vfill}}{}{}%

\begin{document}


\title{A Fourier-based methodology without numerical diffusion for conducting dye simulations and particle residence time calculations}


\author[USC]{Faisal Amlani\fnref{fn1}}
\author[USC]{Heng Wei}
\author[USC,USC2]{Niema M Pahlevan\corref{cor1}}

\cortext[cor1]{Corresponding author: \texttt{pahlevan@usc.edu}}
\fntext[fn1]{\emph{Permanent address:} Universit\'{e} Paris-Saclay, CentraleSup\'{e}lec, ENS Paris-Saclay, CNRS, LMPS - Laboratoire de M\'{e}canique Paris-Saclay, 91190, Gif-sur-Yvette, France}

\address[USC]{Department of Aerospace and Mechanical Engineering, University of Southern California, Los Angeles, USA}
\address[USC2]{School of Medicine, University of Southern California, Los Angeles, USA}
\date{Draft updated: \today}


\date{\today}

\begin{abstract}
  Dye experimentation is a widely used method in experimental fluid mechanics for flow analysis or for the study of the transport of particles within a fluid. This technique is particularly useful in biomedical diagnostic applications ranging from hemodynamic analysis of cardiovascular systems to ocular circulation. However, simulating dyes governed by convection-diffusion partial differential equations (PDEs) can also be a useful post-processing analysis approach for computational fluid dynamics (CFD) applications. Such simulations can be used to identify the relative significance of different spatial subregions in particular time intervals of interest in an unsteady flow field. Additionally, dye evolution is closely related to non-discrete particle residence time (PRT) calculations that are governed by similar PDEs. This contribution introduces a pseudo-spectral method based on Fourier continuation (FC) for conducting dye simulations and non-discrete particle residence time calculations without numerical diffusion errors. Convergence and error analyses are performed with both manufactured and analytical solutions. The methodology is applied to three distinct physical/physiological cases: 1) flow over a two-dimensional (2D) cavity; 2) pulsatile flow in a simplified partially-grafted aortic dissection model; and 3) non-Newtonian blood flow in a Fontan graft. Although velocity data is provided in this work by numerical simulation, the proposed approach can also be applied to velocity data collected through experimental techniques such as from particle image velocimetry.\\
  

\end{abstract}

\begin{keyword}
{Fourier continuation; hemodynamics dye simulation; particle residence time; mathematical physiology; advection-diffusion; high-order methods}
\end{keyword}
\maketitle

\section{\label{sec:introduction}Introduction}

The tracing of dye injections constitutes a fluid visualization technique for the tracking of flows. Dye experimentation is widely used in experimental fluid mechanics for flow analysis or for the study of the transport of particles within a fluid~\cite{bradshaw2016experimental}. Such a method is particularly useful in biomedical diagnostic applications ranging from blood flow analysis in the cardiovascular system to ocular circulation. Although dye tracing is most commonly known as an experimental approach, it can also serve as a useful {post-processing} analysis tool in computational fluid dynamics (CFD) studies~{\cite{kim2004simulated}}. {Performing computational dye simulations can help distinguish the contributions of various different regions within an unsteady flow field.} For example, dye simulations can be applied to the aorta in order to visualize the time histories of the effects of the pulsatile velocity fields on injected medication as well as to investigate the effects of previous cardiac cycles on the distribution of a drug~{\cite{thromb5,zhang2019lattice,kim2004simulated}}. Mathematically speaking, dye simulations are also related to particle residence time (PRT) calculations, both of which can be formulated as convection-diffusion problems~\cite{patankar2018numerical,jozsa,kim2004simulated}.  PRT is an important metric for a variety of applications including those found in environmental fluid dynamics~\cite{maxwell2019exploring,shen2004calculating}, biological flows~\cite{moran1992short}, and hemodynamics~{\cite{reza2019critical,marsden1,marsden2}}. In particular, PRT is now a well-accepted biomarker for cardiovascular diseases (e.g., stroke, myocardial infarction and embolism) since it is inextricably linked to thrombus (blood clot) formation~\cite{thromb1,thromb2,thromb3,thromb4,thromb5}.

While significant effort has been made for the development of various PRT methods (e.g., discrete/non-discrete formulations~{\cite{jozsa,kim2004simulated,marsden1,marsden2}}, Eulerian/Lagrangian formulations~{\cite{maxwell2019exploring,thromb5,reza2019critical}}), little effort has been put towards conducting dye simulations with non-zero diffusion coefficients. Furthermore, to the best of the authors' knowledge, there is currently no numerical methodology in use for dye simulations that incur effectively no numerical diffusion errors {(amplitude errors) or numerical dispersion errors (phase errors)}.  This manuscript introduces a new pseudo-spectral method based on a Fourier continuation (FC) approach for solving the two- and three-dimensional (2D, 3D) convection-diffusion partial differential equation (PDE) systems that govern dye movement and particle residence time evolution without such numerical diffusion errors.

FC-based methods broaden the applicability of Fourier-based spectral methods by enabling Fourier series representations of non-periodic functions while avoiding the well-known ``Gibb's phenomenon." PDE solvers based on such an approach maintain the desirable qualities of spectrally-accurate numerical solvers for time-dependent systems: they can produce accurate solutions by means of relatively coarse discretizations and, importantly, they carry minimal numerical diffusion or dispersion errors (manifesting as artificial amplitude decay or period elongation)---{provided that the corresponding temporal integrator is of sufficiently high-order~\cite{reviewerc}}. FC-based PDE solvers have been successfully constructed for a variety of physical equations including classical wave equations~\cite{lyonbrunoII,brunoprieto}, non-linear Burgers systems~\cite{bruno_jimenez}, Euler equations~\cite{shz,amlanibhat}, Navier-Stokes equations~\cite{albinbruno, cubillos,fontanabruno}, radiative transfer equations~\cite{gaggiolibruno}, Navier elastodynamics equations~\cite{amlanibruno,amlanicarlos,amlanithesis} and 1D fluid-structure hemodynamics equations~\cite{amlanipahlevan}. The numerical methodology introduced in this work {presents} a fully-realized 3D convection-diffusion solver based on this Fast Fourier Transform (FFT)-speed Fourier continuation approximation procedure for both Dirichlet and Neumann boundary conditions.

Convergence and error analyses are presented by considering both the method of manufactured solutions as well as analytical solutions (i.e., based on Gaussians). The performance of the proposed FC methodology is also demonstrated through three physical/physiological applications: 1) flow over a 2D cavity; 2) pulsatile flow in a simplified partially-grafted aortic dissection model; and 3) non-Newtonian blood flow in a Fontan graft.

\section{\label{sec:prtsolver}A high-order method for dye simulations and non-discrete residence time calculations}

This section introduces a new high-order, pseudo-spectral numerical method for dye simulations and non-discrete PRT calculations governed by forced and unforced convection-diffusion equations with Dirichlet or Neumann boundary conditions. The methodology produces solutions based on an FFT-speed Fourier continuation approach~\cite{lyonbrunoI,albinbruno,amlanibruno,amlanipahlevan} for the accurate trigonometric interpolation of non-periodic functions, ultimately enabling high-order spatial accuracy with coarse discretizations using mild {Courant-Friedrichs-Lewy (CFL)} constraints on time integration. Most importantly, such a (spectral) methodology has been demonstrated to faithfully preserve the dispersion/diffusion characteristics of the underlying continuous problems (errors do not compound over distance and time, which may not be true for finite difference or finite element-based methods~\cite{kalinowska2007,sankar1998,mullen,amlanipahlevan,amlanibruno}). These properties are particularly favorable for the accurate assessment of the evolution of an injected dye or drug that may circulate over long times in physiological blood flow configurations (whose corresponding velocity fields, used as input here, are provided by any appropriate fluid-structure solver, e.g., the immersed boundary-lattice Boltzmann method described later{)}.

\subsection{\label{sec:governing} Governing equations}

Given an (incompressible) fluid velocity field $\bv(\bx,t) \in\mathbb{R}^{n}$ on a domain $\bx\in\Omega \subset \mathbb{R}^{n},~n=1,2~\text{or}~3$ with fluid boundary $\partial \Omega\subset \mathbb{R}^{n},~n=1,2~\text{or}~3$, the corresponding evolution of the concentration of a dye~\cite{kim2004simulated, marsden1} or of a non-discrete blood flow particle residence time formulation~\cite{jozsa,reza2019critical,marsden2} can be generally governed by the convection-diffusion system as
\begin{equation}\label{eq:governing}
  \begin{cases}
    \displaystyle\pd{\phi}{t}(\bx,t) + \bv(\bx,t)\cdot \nabla \phi(\bx,t) - \nabla\cdot D(\bx,t) \nabla \phi(\bx,t)= h(\bx,t), & \bx\in\Omega,~t>0,\\
    {\phi(\bx,0) = \phi_0(\bx),} & {\bx\in\Omega,~t=0,}\\
\nabla \phi(\bx,t) \cdot \bn = h_N(\bx,t), & \bx \in \partial \Omega_N,~t\geq0,\\
        \phi(\bx,t) = h_D(\bx,t), & \bx \in \partial \Omega_D,~t\geq0,
  \end{cases}
\end{equation}
where $\bx=(x_1,x_2,x_3)^T$; $\phi(\bx,t)$ is the unknown dye concentration or residence time (as explained in what follows);  {$\phi_0(\bx)$ its initial condition (usually taken to be zero for the cases herein);} $D(\bx,t)\in\mathbb{R}$ is the diffusion coefficient; $h,h_N$ and $h_D \in \mathbb{R}$ are generic forcing functions defined, respectively, on the domain $\Omega$, on the Neumann boundaries $\partial \Omega_N$ ($\bn\in\mathbb{R}^{n}$ is the outward unit normal) and on the Dirichlet boundaries $\partial \Omega_D$. This most general formulation of a convection-diffusion system is treated by the FC solver introduced in the next section; the dye and particle residence time models of interest are merely particular choices of boundaries and forcing functions (described in what follows).

\subsubsection{Dye simulation problem formulation}
Defining $\partial\Omega_D$ to be the flow inlet boundary (where a Dirichlet condition is imposed) and $\partial\Omega_N$ all other boundaries (where a Neumann condition is imposed), the evolution of {a dye} concentration $c(\bx,t) := \phi$, {introduced by an ``injection" along a subset $\Gamma_\text{dye} \subset \partial\Omega_D$ of the inlet boundary, is governed by}~\cite{kim2004simulated, marsden1}
\begin{equation}\label{eq:governingdye}
  \begin{cases}
    \displaystyle\pd{c}{t}(\bx,t) + \bv(\bx,t)\cdot \nabla c(\bx,t) - \nabla\cdot D(\bx,t) \nabla c(\bx,t)= 0, & \bx\in\Omega,~t>0,\\
    {c(\bx,0) = 0,} & {\bx\in\Omega,~t=0,}\\
\nabla c(\bx,t) \cdot \bn = 0, & \bx \in \partial \Omega_N,\\
        c(\bx,t) = 1, & \bx \in \Gamma_\text{dye} \subset \partial \Omega_D,\\
          c(\bx,t) = 0, & \bx \in \partial \Omega_D\backslash  \Gamma_\text{dye},\\
  \end{cases}
\end{equation}
where the Dirichlet value of unity on $\Gamma_\text{dye}$ indicates points at which a concentration is injected (e.g., at a tip of a catheter rather than the entire inlet).

\subsubsection{Non-discrete particle residence time problem formulation}
The convection-diffusion system of Equation~\eqref{eq:governing} also enables a ``non-discrete" (post-processing) formulation for PRT calculations~\cite{jozsa,reza2019critical,marsden2} of a given flow velocity field $\bv(\bx,t)$ (which can be provided either by numerical simulation or experimentation such as through particle image velocimetry~\cite{willert}). Again, defining $\partial\Omega_D$ to be the flow inlet boundary and $\partial\Omega_N$ all other boundaries, the particle residence time $\tau(\bx,t) := \phi$ (given in units of seconds) is governed by Equation~\eqref{eq:governing} with a right-hand-side and null boundary conditions~\cite{jozsa}, i.e.,
\begin{equation}\label{eq:governingprt}
  \begin{cases}
    \displaystyle\pd{\tau}{t}(\bx,t) + \bv(\bx,t)\cdot \nabla \tau(\bx,t) - \nabla\cdot D(\bx,t) \nabla \tau(\bx,t)= 1, & \bx\in\Omega,~t>0,\\
    {\tau(\bx,0) = 0,} & {\bx\in\Omega,~t=0,}\\
\nabla \tau(\bx,t) \cdot \bn = 0, & \bx \in \partial \Omega_N\\
          \tau(\bx,t) = 0, & \bx \in \partial \Omega_D.\\
  \end{cases}
\end{equation}
 The Dirichlet value of zero at the inlet follows from the particular configurations of interest (i.e., hemodynamics applications): since fluid flows downstream from the inlet, particles along the inlet have not spent time inside the domain $\Omega.$ For some PRT solvers used for hemodynamics applications~\cite{marsden1,marsden2}, the diffusion ${D}(\bx,t)$ is often taken to be zero (no such assumption is required for the proposed methodology of this paper). For those that have considered mass diffusivity~\cite{kim2004simulated,reza2019critical}, {no analysis of numerical diffusion errors has been presented} (finite difference or finite element methods are well-known to suffer from such errors~\cite{kalinowska2007,sankar1998,mullen,amlanipahlevan,amlanibruno}, including for convection-diffusion equations~\cite{sankar1998}).

\subsection{\label{sec:FC} A pseudo-spectral Fourier continuation methodology without numerical diffusion or dispersion errors}

\subsubsection{Spatial discretization}
For the numerical treatment of the spatial variables and directional derivatives of Equation~\eqref{eq:governing}, one can consider
discrete point values $c(x_i)$ of a given smooth function $c(x) : [0,1]
\rightarrow \mathbb{R}$ (defined on the unit interval without loss of generality) through a uniform discretization of size $N$, i.e., $x_i = i \Delta x,~i=0,\dots,N-1,~\Delta x = 1/(N-1)$. The FC method constructs a fast-convergent Fourier series (or trigonometric interpolant) $c_{\text{cont}} : [0,b] \rightarrow \mathbb{R}$ (where $b$ is only slightly larger than $1$) that is given by
\begin{equation}\label{eq:fcseries} c_{\text{cont}} = \displaystyle\sum_{k=-W}^{W} a_k e^{\frac{2\pi i k x}{b}} \quad \text{s.t.}\quad c_{\text{cont}}(x_i) = c(x_i),~i=0,...,N-1,\end{equation}
where $W = (N+N_\text{cont})/2$ is the corresponding bandwidth for an $N_\text{cont}$-sized truncated Fourier series (here, $N_\text{cont}$ is defined as the number of discrete points added in the interval $[1,b]$, i.e., $b=(N+N_\text{cont})\Delta x$). The continued (or ``extended") function $c_{\text{cont}}$ renders the original $c(x)$ periodic, discretely approximating $c$ to very high-order in $[0,1]$ but demonstrating periodicity on the slightly larger
$[0,b]$.  Directional spatial derivatives on a dimension $x$ of $c(\bx,t) = c(x,y,z,t)$ in Equation~\eqref{eq:governing} can then be calculated by exact termwise differentiation of the trigonometric series given by Equation~\eqref{eq:fcseries} as
\begin{equation}\label{eq:termwise}
\pd{c_{\text{cont}}}{x}(x) = \displaystyle\sum_{k=-W}^{W} \left(\frac{2\pi i k}{b}\right) a_k e^{\frac{2\pi i k x}{b}}.
\end{equation}
By restricting the domain of $\partial c_{\text{cont}}/\partial x$ to the original unit interval, the numerical derivatives of $c$ are hence approximated to high-order. The overall errors are found only in the production of Equation~\eqref{eq:fcseries}, from where the calculation of the derivative {in}~Equation~\eqref{eq:termwise} is easily facilitated by {an FFT}. This overall idea can also be easily extended to any general interval $[x_0,x_N]$ via affine transformations~\cite{lyonbrunoI}.

In {the} most simple treatment~\cite{brunohan,boyd}, the coefficients $a_k$ of Equation~\eqref{eq:fcseries} can be found through the solution of a least squares system given by
\begin{equation}\label{eq:leastsquares}
\displaystyle\min_{a_k} \sum_{i=0}^{N-1} \left|c_{\text{cont}}(x_i) - c(x_i)\right|^2,
\end{equation}
which can be solved through a Singular Value Decomposition (SVD). However, for a  PDE that evolves in time such as the convection-diffusion equations for dye simulations, time integration can incur significant computational expense (where the SVDs must be applied at each timestep in multiple spatial dimensions). For tractability, this work employs an accelerated continuation technique known as FC(Gram)~\cite{lyonbrunoI,lyonbrunoII, albinbruno}. Relying on the use of very small (e.g., five) numbers of function values near the left ($x=0$) and right ($x=1$) endpoints, the FC(Gram) algorithm projects such values onto an interpolating (Gram) polynomial basis. Fourier continuations are precomputed for each (significantly reduced) basis vector through solving the corresponding least squares formulation similar to Equation~\eqref{eq:leastsquares} by a high-precision or symbolic SVD. That is, via a subset of given function values on small numbers $d_\ell$ and $d_r$ of matching points
$\{x_0,..,x_{d_\ell-1}\}$ and $\{x_{N-d_r},...,x_{N-1}\}$ at the leftmost and rightmost sub-intervals of $[0,1]$, one can produce the discrete periodic extension of size $N_\text{cont}$ by projecting these end values onto a Gram polynomial basis up to degree $d_\ell-1$ (or $d_r-1$). This constructs an interpolating polynomial whose Fourier continuations can be precomputed, where orthogonality is enforced by the scalar product naturally defined by the discretization points. This ultimately forms a (precomputed) ``basis" of continued functions that can represent  any function $c$  and provide a smooth transition from $c(x=0)$ to $c(x=1)$ over the slightly larger interval $[0,b]$.

The solver introduced in this work invokes a particular ``blend-to-zero" variation of FC(Gram)~\cite{albinbruno,amlanibruno,amlanipahlevan}. Instead of continuing to the opposite endpoint of the function, the blending algorithm transitions the matching points on the left and the right smoothly to values of zero at the  discrete points $\{x_{-N_{\text{cont}}-d_r},...,x_{-N_{\text{cont}}-1}\}$ and $\{x_{N+N_{\text{cont}}},...,x_{N+N_{\text{cont}}+d_\ell-1}\}$. The sum of the two leftward and rightward continuations constructs the complete (discretely-periodic) Fourier continuation. Each blending transition is \emph{precomputed} on each element of the Gram polynomial basis (that interpolates the handful of endpoint values of $c$) by solving the corresponding SVD minimization. At the left, for example, this Gram basis can be constructed by orthogonalizing a Vandermonde matrix
 \begin{equation}\label{eq:vander}
 V = \begin{pmatrix}
 1 & x_0 & (x_0)^2 & ... &(x_0)^{d_\ell-1}\\
 1 & x_1 & (x_1)^2 & ... &(x_1)^{d_\ell-1}\\
 \vdots & \vdots & \vdots & \vdots &\vdots\\
 1 & x_{d-1} & (x_{d-1})^2 & ... &(x_{d_\ell-1})^{d_\ell-1}\\
 \end{pmatrix}
\end{equation}
  of point values at the discrete points $\{x_0,...,x_{d_\ell-1}\}$ of the monomials $\{x^0,x^1,..,x^{d_\ell-1}\}$ (a substitution of $1-x$ gives the {equivalent} representation for the rightward matching problem). An orthogonalization can then be performed by employing Gram-Schmidt ($V=QR$) in high-precision (see~\cite{albinbruno,amlanibruno} for details). This ultimately produces projection operators $Q_\ell$ and $Q_r$ (on the left and the right, respectively) in order to determine the coefficients in the polynomial bases for interpolating the matching endpoints of the function $c$.  Defining vectors of such sets of points for the left and right by
 \begin{equation}\label{eq:flfr}
 \vc_\ell = \left(c(x_0), c(x_1), ..., c(x_{d_\ell-1})\right)^T, \quad  \vc_r = \left( c(x_{N-d_r}), c(x_{N-d_r+1}),..., c(x_{N-1})\right)^T,
 \end{equation}
   the overall FC operation can hence be expressed as
\begin{equation}\label{eq:FCoperator}
\vc_{\text{cont}} = \begin{bmatrix} \vc \\ A_\ell Q_\ell^T \vc_\ell + A_r Q_r^T \vc_r\end{bmatrix} = \begin{bmatrix} \vc \\ P_\ell \vc_\ell + P_r \vc_r\end{bmatrix}
\end{equation}
where $\mathbf{c} = (c(x_0),\dots,c(x_{N-1}))^T$ contains the discrete point values of $c$; $\vc_{\text{cont}}$ is a vector of the $N+N_{\text{cont}}$ discretely-periodic function values; $I$
is the identity matrix; and $A_\ell, A_r$ contain the
corresponding $N_\text{cont}$ values that blend, to zero, the left and the right Gram basis vectors. The columns of $Q_\ell$, $Q_r$ (containing the $d_\ell, d_r$
point values of each element of the corresponding Gram basis) differ only by column-ordering for $d_\ell=d_r$ (and hence $A_\ell, A_r$ by row). Further technical details on the blend-to-zero FC(Gram) algorithm have been provided previously~\cite{lyonbrunoI,albinbruno,amlanibruno}.

For the generalized system given by Equation~\eqref{eq:governing}, {Neumann} (derivative) conditions are often invoked for residence time and dye simulations at boundaries away from the inlet. The methodology outlined above requires function values at the absolute left and right physical endpoints (as part of the matching sets of $d_\ell, d_r$ points), which is naturally satisfied by a Dirichlet boundary-value problem. If a Neumann boundary value is specified (for example, at the right endpoint as $c'(x_0)$), it is necessary to
match a given derivative value in addition to given function values contained in the rest of the matching interval (i.e., $\{x_1,...,x_{d_\ell-1}\}$). This can be accomplished by following a similar procedure to the classic FC(Gram) outlined above, as described previously~\cite{amlanibruno}: a modified polynomial interpolant can be introduced to
match the derivative value $c'(x_{0})$ at the endpoint $x_{0}$ 
as well as the function values at $x_1,...,x_{d_\ell-1}$. Such an
interpolant can be obtained by orthonormalizing, instead of the columns of Equation~\eqref{eq:vander}, the corresponding columns of a
modified Vandermonde matrix given by
\begin{equation}\label{eq:polyderiv}
V_{\mathrm{mod}} = \begin{pmatrix}
0 & 1 & 2 x_{0} & ... &(d_\ell-1)(x_{0})^{d_\ell-2}\\
1 & x_{1} & (x_{1})^2 & ... &(x_{1})^{d_\ell-1}\\
\vdots & \vdots & \vdots & \vdots &\vdots\\
1 & x_{d_\ell-1} & (x_{d_\ell-1})^2 & ... &(x_{d_\ell-1})^{d_\ell-1}
\end{pmatrix}
\end{equation}
by means of a QR-decomposition that similarly produces the operators
\begin{equation}\label{eq:QRder}
V_{\mathrm{mod}} = Q_{\mathrm{mod}}R_{\mathrm{mod}}.
\end{equation}

In order to use the same precomputed Fourier continuation basis data produced for the Dirichlet case, a reconstruction of the coefficients from the Neumann case to the original
Gram polynomial basis can be simply obtained by replacing the operator $Q_\ell$ (resp. $Q_r$) in Equation \eqref{eq:FCoperator} with a modified operator $\bar{Q}_\ell$
(resp. $\bar{Q}_r$) that directly performs such a (re-)projection.  The new operator can be found by first solving for new coefficients $\ba = \left(a_1,a_2,...,a_{d-1}\right)^T$ via the expression
\begin{equation}\label{eq:polyderivcoeff}
V_{\mathrm{mod}}\ba= \left(c_0, c_1, ..., c_{d-2}, c'(x_{d-1})\right)^T
\end{equation}
which, from the QR-decomposition in Equation~\eqref{eq:QRder}, can be solved as
\begin{equation}\label{eq:coefficients}
\ba = R_{\mathrm{mod}}^{-1} Q_{\mathrm{mod}}^{T}
\left(c_0, c_1,...,c_{d-2},c'(x_{d-1})\right)^T.
\end{equation}
From the original decomposition of the original Vandermonde matrix $V=QR$ of Equation~\eqref{eq:vander}, the corresponding coefficients are similarly given by
\begin{equation}\label{eq:coefficients2}
V\ba = QR\ba\left( c_0, c_1,...,c_{d-2},c_{d-1}\right)^T.
\end{equation}
Substitution of Equation~\eqref{eq:coefficients} into Equation~\eqref{eq:coefficients2} yields the expression
\begin{equation*}
QRR_{\mathrm{mod}}^{-1} Q_{\mathrm{mod}}^{T}
\left( c_0,c_1,...,c_{d-2},c'(x_{d-1})\right)^T   =  \left( c_0, c_1,...,c_{d-2},c_{d-1}\right)^T.
\end{equation*}
Defining $\bar{Q} = (RR_{\mathrm{mod}}^{-1} Q_{\mathrm{mod}}^{T})^T$ further gives
\begin{equation}\label{eq:polysystemder}
\bar{Q}^T
\left(c_0, c_1, ..., c_{d-2}, c'(x_{d-1})\right)^T  = Q^T\left( c_0, c_1,...,c_{d-2},c_{d-1}\right)^T.
\end{equation}
Hence the new block matrix form of the continuation with the modified operators are by \begin{equation}\label{eq:FCoperatormod}
\vc_{\text{cont}} = \begin{bmatrix} \vc \\ A_\ell \bar{Q}_\ell^T \bar{\vc}_\ell + A_r Q_r^T \vc_r\end{bmatrix} = \begin{bmatrix} \vc \\ P_\ell \bar{\vc}_\ell + P_r \vc_r, \end{bmatrix}
\end{equation}
where $\bar{c}_\ell = (c'(x_{0}),c_1,...,c_{d_\ell-1})^T,~\vc_r =
(c_{N-d_r},...,c_{N-1})^T$ and where $A_\ell, A_r$ are
the same original blend-to-zero operators from the Dirichlet case. Such a formulation enables mixing and matching of left-right boundary conditions (e.g., Neumann-Dirichlet, Neumann-Neumann), where a similar procedure can be performed to find the corresponding projection operator for Neumann boundary conditions at the right endpoint.

\begin{figure}
  \begin{center}
\includegraphics[width=.485\textwidth]{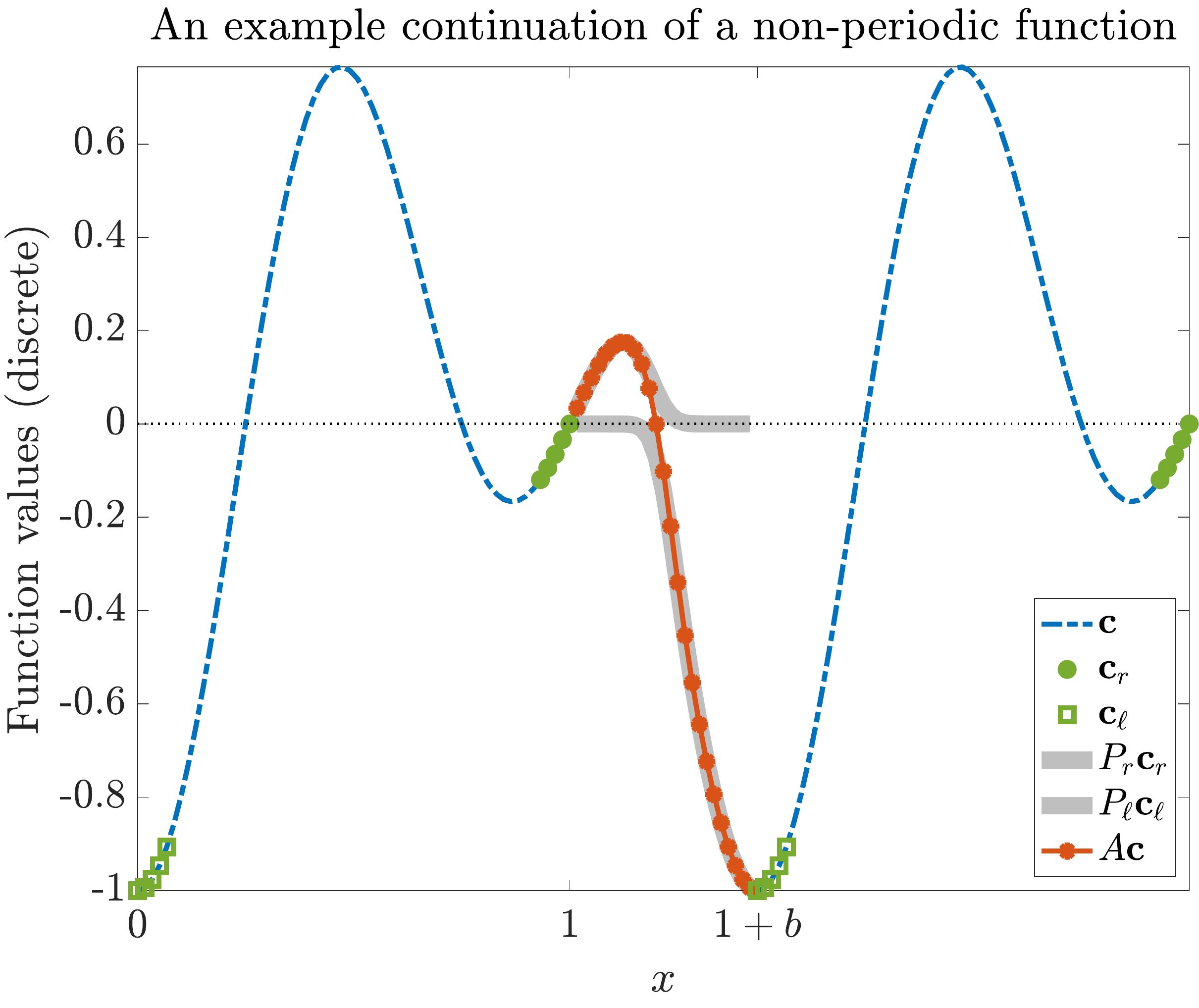}
\end{center}
\caption{\label{fig:fig_1}An example Fourier continuation of the non-periodic function $c(x) = \left(1-x^2\right)\sin\left(2\pi x-\pi/2\right), x\in[0,1]$. The operators $P_\ell, P_r$ perform the blending-to-zero via projection onto a precomputed FC basis. The new function values $\mathbf{c}_\mathrm{cont}$ are periodic (in a discrete sense) on the interval $[0,b]$.  }
\end{figure}

In summary, the FC(Gram) algorithm appends $N_{\text{cont}}$ values to a given discretized function in order to form a periodic extension in $[1,b]$ that transitions smoothly
from $c(x_{N-1})$ back to $c(x_0)$ (or, in the case of known Neumann boundaries, back to, e.g., $c'(x_0)$). An illustrative example is provided in Figure~\ref{fig:fig_1}, where a non-periodic function is originally  discretized on $[0,1]$ and then translated by a distance $N_\text{cont}\Delta x$ with the subsequent interval filled-in by the sum of leftward and rightward blend-to-zero continuations. The resulting continued vector $\bc_{\text{cont}}$ can be interpreted as a set of discrete values of a smooth and periodic function that can be approximated to high-order via FFT on an interval
of size $(N+N_{\text{cont}})\Delta x$ (and hence accurate termwise derivative representations for Equation~\eqref{eq:governing}). As has been suggested previously~\cite{amlanibruno,amlanipahlevan}, a small number of $d_\ell, d_r = 5$ matching points with a periodic extension comprised of $N_\text{cont} = 25$ points are used for all simulations presented hereafter. {Such values have been previously demonstrated to provide a good balance of accuracy---here, fifth-order in space---and the computational cost associated with the extension procedure~\cite{albinbruno,amlanipahlevan,amlanibruno,lyonbrunoI,bruno_jimenez, amlanicarlos, gaggiolibruno}. They are determined by numerical experiments, which additionally suggest that higher orders may lead to numerical instability from physical boundaries (observations which are consistent with those previously reported~\cite{albinbruno,lyonbrunoI})}. The matrices $A_\ell, A_r, Q_\ell, Q_r, \bar{Q}_\ell$ and $\bar{Q}_r$ are computed only once (before numerically solving the governing system) and stored in file for use each time the FC procedure is invoked. {In summary, the FC method discretely extends a non-periodic function so that it is periodic and amenable to high accuracy FFTs, using a precomputation procedure that appends a fixed number of points to the original discretized domain (see Figure~\ref{fig:fig_1} for a visualization).} {These precomputations, coupled with the aforementioned choices of parameters, lead to a continuation process whose computational cost amounts to a fixed $25\times5$ by $5\times1$ matrix-vector multiplication (see Section~\ref{sec:performance} for a cost analysis of the FC procedure).}

\subsubsection{Temporal integration}

An explicit Adams-Bashforth scheme of order four (AB4) is employed for marching forward-in-time, similarly to other FC-based solvers~\cite{albinbruno,amlanibruno,amlanipahlevan,amlanibhat}. Other explicit timesteppers that provide adequate regions of stability~\cite{bashforth,rk4musa} can also be employed, although timesteppers such as the fourth-order Runge-Kutta method (RK4) entails four evaluations of the right-hand-side $F(\bx,t)$ that is given  (from Equation~\eqref{eq:governing}) by
\begin{equation}\label{eq:governingrhs}
  \displaystyle\pd{c}{t}(\bx,t) = - \bv(\bx,t)\cdot \nabla c(\bx,t) + \nabla\cdot D(\bx,t) \nabla c(\bx,t) + h(\bx,t) = F(\bx,t).
\end{equation}
Multiple evaluations of $F$ at each timestep for RK4 can become rather costly for 2D and 3D simulations (where the spatial derivatives must be computed at each evaluation), and hence an AB4 method is used instead {(see Section~\ref{sec:performance} for a numerical comparison of cost and accuracy)}. For a uniform timestep $\Delta t$ and a given time $t$, the corresponding equation for integrating to $t+\Delta t$ is given by
\begin{equation}\label{eq:systemforAB4}
    \displaystyle\pd{c}{t}(\bx,t+\Delta t) = c(\bx,t) + \frac{\Delta t}{24} \left(55 F(\bx, t)-59 F(\bx, t-\Delta t)+37 F(\bx, t-2\Delta t) - 9 F(\bx, t-3\Delta t)\right).
  \end{equation}
Dirichlet boundary conditions can then imposed after each timestep $t\to t+\Delta t$ by direct injection, and Neumann boundary conditions are implemented naturally when the modified Fourier continuation projection operator is invoked to compute a spatial derivative at $t$.  For all simulations in this paper, a timestep of
\begin{equation}\label{eq:CFL}
  \Delta t \leq \frac{\sqrt{2}}{2}\frac{1}{\max\|\bv(\bx,t)\|}\min \Delta x_i
\end{equation}
   has provided absolute stability in all cases, including runs of millions of timesteps. {Equation~\eqref{eq:CFL}  has been determined empirically by numerical experiments, and is consistent with those proposed for other FC-based solvers~\cite{lyonbrunoII,amlanipahlevan,ellingthesis,albinbruno}.} {In addition to such mild CFL constraints, the corresponding Peclet numbers ($D\Delta t /\Delta x^2$) for the cases and parameters considered in this work are adequately small (all less than $0.1$) and well within prescribed bounds to maintain numerical stability for the corresponding linear convection-diffusion problems~\cite{reviewera}.}

Similarly to other spectral solvers, a spatial filter is employed at each timestep above to control the growth of errors in unresolved modes~\cite{albinbruno,amlanibruno,amlanipahlevan}. Applied in Fourier space on a function $g(x)$ with Fourier coefficients $\hat{g}_k$, such a filter is given by
\begin{equation}\label{eq:filter}
  \sum_{k=-\frac N2}^\frac{N}{2} \hat{g}_k \exp(ikx)\quad \longrightarrow\quad \sum_{k=-\frac N2}^\frac{N}{2} \exp\left(-\alpha\left({2k}/{N}\right)^{{2p}}\right) \hat{g}_k \exp(ikx),
\end{equation}
where $p$ is a positive integer that determines the rate of decay of the coefficients, and where the real valued $\alpha>0$ determines the level of
suppression such that the highest-frequency modes are multiplied by $e^{-\alpha}.$ For all numerical in this paper, values of $(p,\alpha) = \left(8,-1/10\times \ln\left(10^{-2}\right)\right)$ are employed similarly to other FC-based solvers (further discussion on filtering can be found elsewhere~\cite{amlanibruno,ellingthesis,amlanipahlevan}), {and do not completely eliminate the highest-frequency modes as in some previously adopted choices~\cite{albinbruno}.}

\subsection{Fluid-solid interaction solver: an immersed boundary-lattice Boltzmann method\label{LBMsec}}

The input fluid velocity field $\bv(\bx,t)$ of Equation~\eqref{eq:governing} can be provided by any suitable flow solver. For the hemodynamics applications of this work, such velocity vectors are produced by a non-Newtonian fluid-structure solver based on a {lattice Boltzmann method (LBM)} for the fluid (which has been proven particularly useful for hemodynamics applications~\cite{randles,ames2020multi,wei2020,wu2017lattice}). This method employs simplified kinetic equations in conjunction with a modified molecular-dynamics approach, where the motions of particles on a regular lattice are enforced by distribution functions that provide mass and momentum conservation. For a position $\bx$ and a time $t$, the distribution functions of this work are governed by the Bhatnagar-Gross-Krook collision model~\cite{bhatnagar} through the expressions
\begin{eqnarray}
\label{eqn:fi}
f_i(\boldsymbol{x}+\boldsymbol{e}_i\Delta t,t+\Delta t)-f_i(\boldsymbol{x},t)=-\frac{1}{T}[f_i(\boldsymbol{x},t)
-f_i^{eq}(\boldsymbol{x},t)]+\Delta t F_i,
\end{eqnarray}
where $f_i(\mathbf{X},t)$ are distribution functions of the particles in phase space ($f_i^{eq}$ is the corresponding equilibrium distribution); $\mathbf{e}_i$ are discrete velocities; $\Delta t$ is the temporal discretization; and $T$ is a (non-dimensionalized) relaxation time that is related to fluid viscosity. For a spatial discretization $\Delta x = \Delta x_i$, the corresponding pressure $P$ and flow velocity $\bv$ are given respectively by
\begin{eqnarray}
\label{eqn:rho}
P=v_s^2\sum_if_i,
\end{eqnarray}
and
\begin{eqnarray}
\label{eqn:v}
\boldsymbol{v}=\frac{1}{\rho}\sum_i\boldsymbol{e}_if_i+\frac{1}{2}\boldsymbol{f}\Delta t,
\end{eqnarray}
where $v_s = \Delta x/(\sqrt{3}\Delta t)$ is the lattice sound speed. In all the case studies presented herein, a so-called D2Q9 stencil is used, yielding $i=0,...,9$ discrete velocity vectors. Further details on the lattice Boltzmann method can be found in literature~\cite{mohamad2011lattice,chen1998lattice}.

For problem formulations including compliant (elastic) walls (e.g., the simplified partially-grafted aortic dissection model described later), a corresponding structure equation can be considered in Lagrangian coordinates as

\begin{equation}
\label{eqn:structure}
\rho_s h \frac{\partial ^2\mathbf{X}}{\partial t^2}=\frac{\partial}{\partial s}
\Bigg[Eh\left(1-\left(\frac{\partial \mathbf{X}}{\partial s}\cdot\frac{\partial \mathbf{X}}{\partial s}\right)^{-1/2}
\right)\frac{\partial \mathbf{X}}{\partial s}-\frac{\partial}{\partial s}\left(EI\frac{\partial ^2\mathbf{X}}{\partial s^2}\right)\Bigg]+\mathbf{F}_{L},
\end{equation}
where $\rho_{s}$ is the density of the solid wall; $h$ is the thickness; $s$ is the Lagrangian coordinate along the solid wall; $\mathbf{X}$ is the position (deformation) of the solid wall; $\mathbf{F}_{L}$ is the Lagrangian force exerted on the solid wall by the fluid; and $Eh$ and $EI$ are the stretching {stiffness} and bending {stiffness}, respectively. The boundary condition of the solid wall at a free end is given by
\begin{equation}
\label{eqn:boundaryconditionfree}
1-\left(\frac{\partial \mathbf{X}}{\partial s}\cdot\frac{\partial \mathbf{X}}{\partial s}\right)^{-1/2}=0,
~~~ \frac{\partial ^2\mathbf{X}}{\partial s^2}=(0,0),~~~ \frac{\partial ^3\mathbf{X}}{\partial s^3}=(0,0),
\end{equation}
and at a fixed end (simply supported) is given by
\begin{equation}
\label{eqn:boundaryconditionfixed}
\mathbf{X}=\mathbf{X}_o,\qquad \frac{\partial\mathbf{X}}{\partial s}=(-1,0).
\end{equation}
Solving the solid equations via a {finite element method (FEM)}, the immersed boundary (IB) method~{\cite{goldstein1993modeling,ames2020multi}} is used to ultimately couple the solid solution with the LBM fluid solver~\cite{huang2007simulation, hua2014dynamics}.
This method has been extensively used to simulate fluid-structure interaction (FSI) problems in cardiovascular biomechanics
~\cite{peskin2002immersed,mittal2005immersed}. The reference quantities used to non-dimensionalize both the fluid and solid formulations are given by the density $\rho$; velocity $U_\infty=Q/A$ (where $A$ is the inlet area); and length $L=D$ ($D$ is the inlet diameter). The subsequent non-dimensional physical parameters are given by the Reynolds number $\mathrm{Re}=\rho U_\infty L/\mu$; the Womersley number $\mathrm{Wo}=\sqrt{\rho \omega L^2/\mu}$; the bending coefficient $K=EI/\rho U_\infty^2L^3$; the tension coefficient $S=Eh/\rho U_\infty^2L$; and the mass ratio of the solid wall to the fluid given by $M=\rho_s h/\rho L$.

\section{\label{sec:performance}Convergence \& error analysis}

\subsection{Method of manufactured solutions for verifying implementation, accuracy \& convergence}

For the proposed FC-based methodology, the method of manufactured solutions~\cite{amlanibruno,amlanipahlevan,roache,roy,vedovoto,mmsraghu, mmsraghu2} (MMS) has been employed in order to verify both its numerical accuracy as well as its code implementation. Such a method postulates a (sufficiently smooth) solution of the convection-diffusion PDE and, upon substitution, incorporates the corresponding right-hand side and boundary conditions as non-trivial forcing terms. The resulting system enables the proposed function
to be an exact solution of a forced convection-diffusion system. For example, one can postulate a solution to $c(\bx,t)$ as
\begin{equation}\label{eq:MMS}
c(\bx,t) = \cos(2\pi f_1(x_1-\kappa_1 t))\cos(2\pi f_2(x_2-\kappa_2 t)),
\end{equation}
which, upon {substitution} into the PDE of Equation~\eqref{eq:governing}, yields a corresponding right-hand-side as
\begin{eqnarray*}\label{eq:rhs}
  h(\bx,t) &=& 2\pi \big[\kappa_1 f_1 \cos(2\pi f_2(x_2 - \kappa_2t))\sin(2\pi f_1 (x_1 - \kappa_1 t))\\
  & +& \kappa_2 f_2\cos(2\pi f_1(x_1 - \kappa_1 t))\sin(2\pi f_2 (x_2 - \kappa_2 t))\big] \\
  & -&2\pi \bv(\bx,t)\cdot\begin{pmatrix}f_1\cos(2\pi f_2(x_2 - \kappa_2 t))\sin(2\pi f_1(x_1 - \kappa_1 t))\\f_2\cos(2\pi f_1(x_1 - \kappa_1 t))\sin(2\pi f_2(x_2 - \kappa_2 t))\end{pmatrix}.
\end{eqnarray*}
{Here, the initial condition of the problem is given by the corresponding exact solution (Equation~\eqref{eq:MMS}) evaluated at $t=0.$}

\begin{figure}
\includegraphics[width=.485\textwidth]{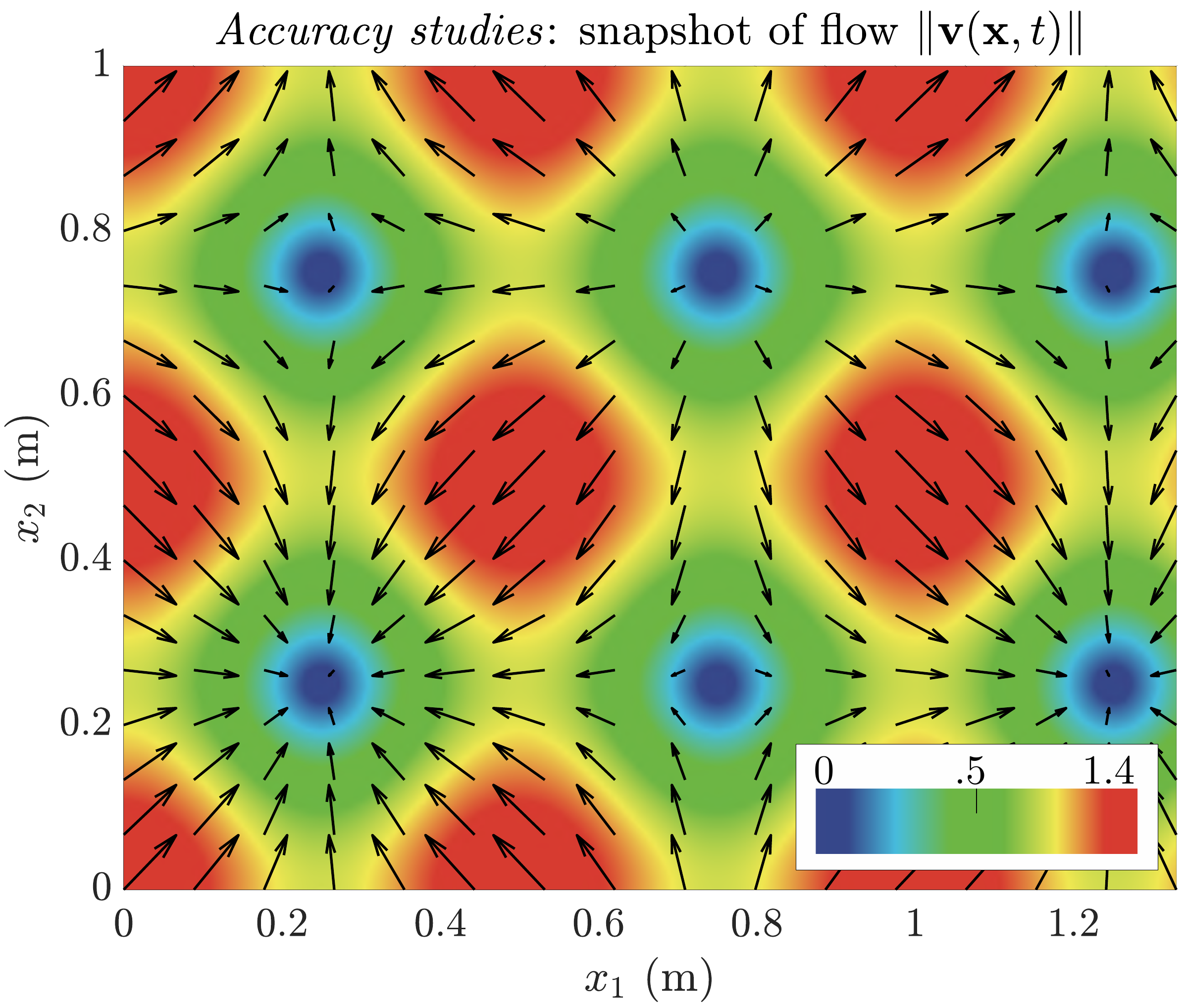}\hfill
\includegraphics[width=.485\textwidth]{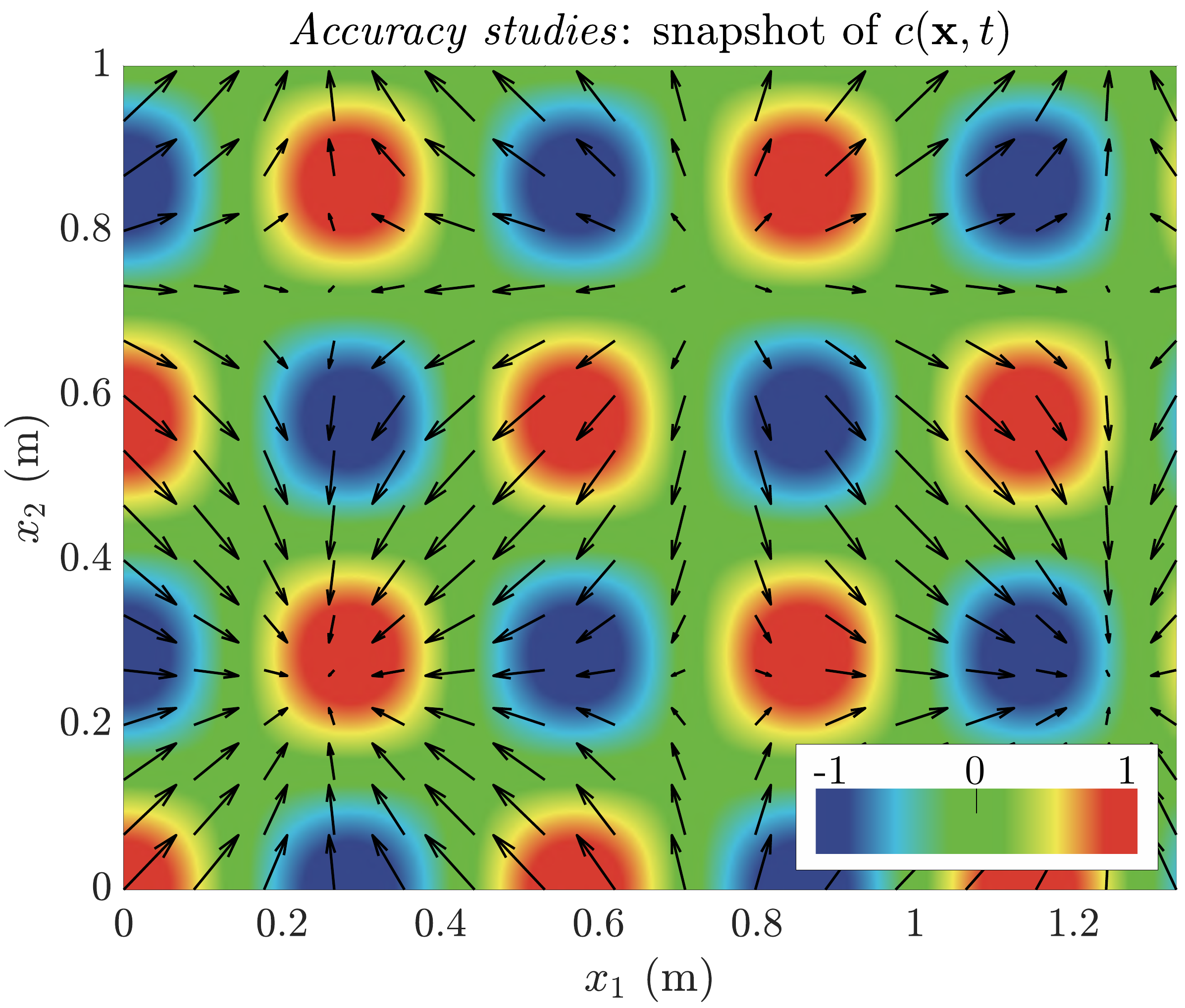}
\caption{\label{fig:fig_2}(Left) Snapshot of the magnitude $\|\bv(\bx,t)\|$ of the assumed flow velocity  and (right) the corresponding FC-produced numerical snapshots (computed using $128\times128$ discretization points) of the manufactured convection-diffusion solution $c(\bx,t)$ corresponding to Equation~\eqref{eq:MMS}. Velocity stream directions are overlaid in black. }
\end{figure}

Assuming a manufactured closed-form velocity field given by
$$\bv = (\cos(2\pi x) \cos(f_v 2\pi t),\sin(2\pi y) \cos(f_v 2\pi t))^T,$$
and the parameters $f_v = 12$ Hz, $f_1=f_2 = 1.75$ Hz, $\kappa_1 = -.25$ m/s, $\kappa_2 = .25$ m/s and $D=1\times10^{-2}$ m$^2$/s, Figure~\ref{fig:fig_2} presents a snapshot of the corresponding numerical solution given by {Equation}~\eqref{eq:MMS} at an arbitrarily chosen moment in time in a domain defined by $\Omega =  [0,1.33]\times[0,1.00]$.

\begin{figure}
\includegraphics[width=.485\textwidth]{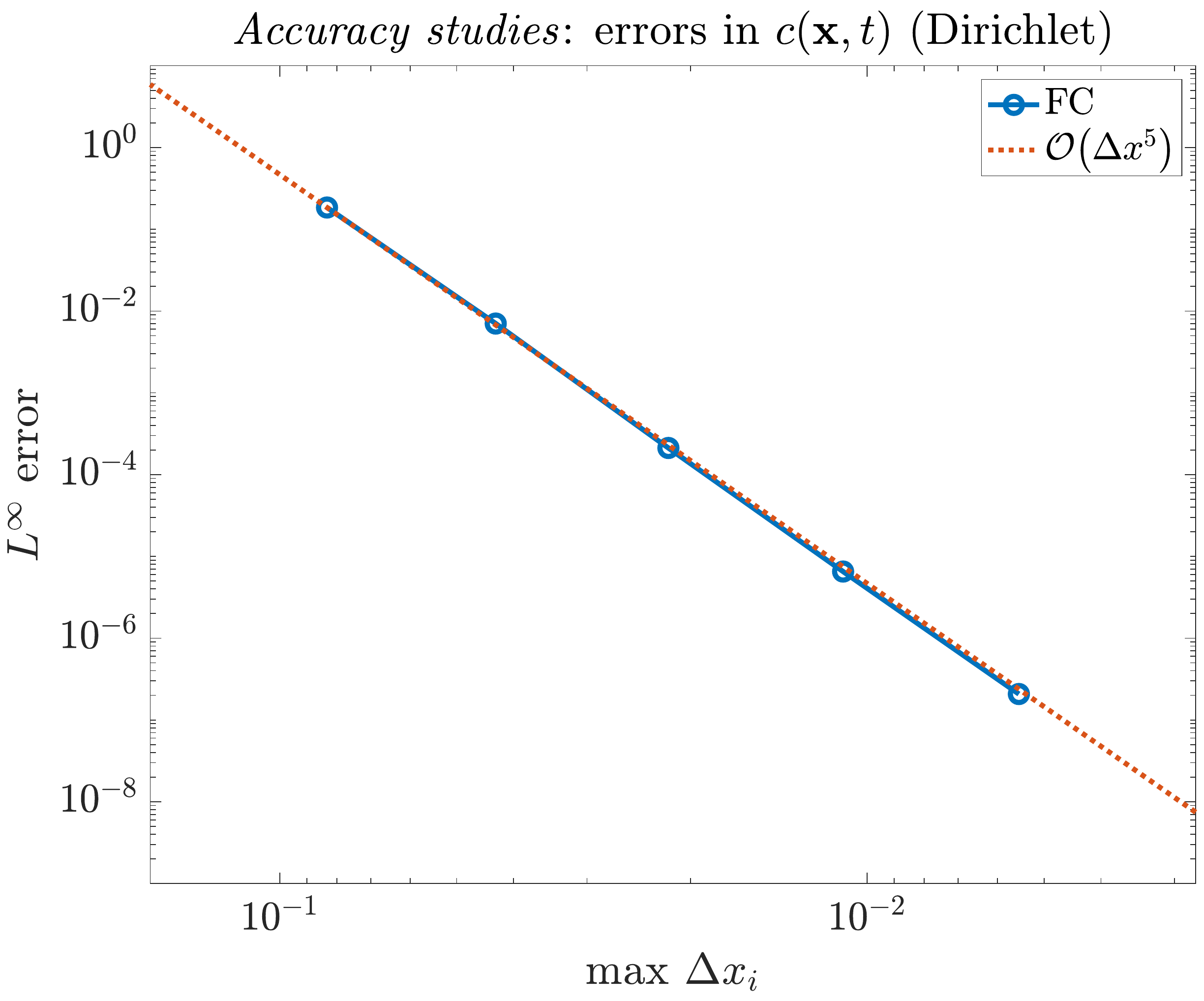}\hfill
\includegraphics[width=.485\textwidth]{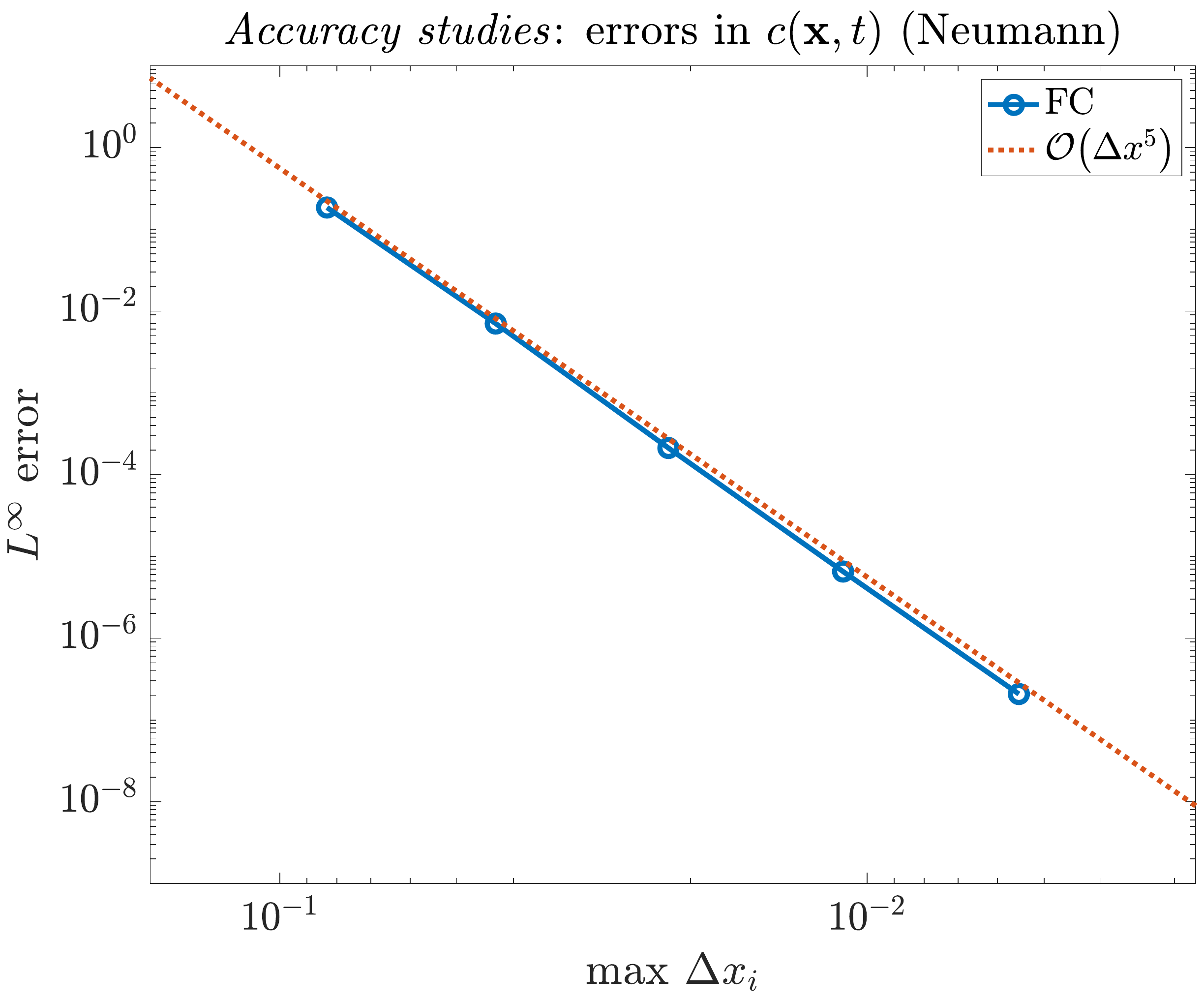}
\caption{\label{fig:fig_3}The maximum ($L^{\infty}$) errors over all space and time after $5000$ timesteps for the manufactured right hand side incorporating (left) Dirichlet and (right) Neumann boundary conditions. Errors are calculated by comparison with the exact solution (Equation~\eqref{eq:MMS}) and with a fine-enough timestep so that errors are dominated by spatial discretization. The overlaid slopes demonstrate fifth-order convergence as expected from employing $d_\ell, d_r = 5$ matching points to construct Fourier continuations.}
\end{figure}

 Applying the FC solver on a series of discretizations corresponding to integer divisions of $\Delta x_1 = \Delta x_2 = 1/32$ (where all timesteps are chosen small enough so that errors are dominated by spatial discretizations), Figure~\ref{fig:fig_3} presents the maximum absolute $L^\infty$ errors, over all space and all time (up to $5000$ timesteps), for both Dirichlet boundary conditions (Figure~\ref{fig:fig_3}, left) and Neumann boundary conditions (Figure~\ref{fig:fig_3},  right). The overlaid slopes in these plots illustrate the expected fifth-order accuracy of the operators from the FC parameters employed in this work.

{Table~\ref{tab:table1} additionally presents a comparison after $100000$ timesteps of errors and computational times between multi-stage AB4 and RK4 timesteppers for the above manufactured solution and velocity formulation corresponding to $f_v = 4$ Hz, $f_1=f_2 = 1.75$ Hz, $\kappa_1 = -.25$ m/s, $\kappa_2 = .25$ m/s, $D=1\times10^{-2}$ m$^2$/s, and Dirichlet boundary conditions. As can be observed, errors are similar between both time integration schemes. However, as discussed earlier and evident in the table, the additional evaluations of the right-hand-side for RK4 lead to an overall computational cost that is four times larger than the cost associated with employing AB4, supporting the adoption of the latter in this work. For a detailed analysis of multi-step time integration schemes for advection-like problems, the reader is referred to~\cite{reviewerd,reviewere}. Table~\ref{tab:table1} additionally presents the associated cost in seconds of performing the continuation procedure as well as its corresponding proportion (in percentage) of the overall computational cost of the solutions. Results indicate that the cost is no more than $12\%$ of the overall time for AB4 (using the coarsest discretization), which subsequently decreases with increasing degrees-of-freedom (to around four percent for the finest discretization). This indicates that the additional burden of the Fourier continuation procedure to affect a periodic extension is minimal.}

{
\begin{table*}

  \addtolength{\tabcolsep}{-0.2em}
  \caption{\label{tab:table1}Comparison of maximum $L^\infty$ errors as well as computational times (in seconds) between using AB4 and RK4 time integrators in an FC-based solver. The manufactured solution and velocity parameters correspond to $f_v = 4$ Hz, $f_1=f_2 = 1.75$ Hz, $\kappa_1 = -.25$ m/s, $\kappa_2 = .25$ m/s, $D=1\times10^{-2}$ m$^2$/s, and Dirichlet boundary conditions. Errors are computed after $100000$ timesteps by comparison with the exact solution and with a fine-enough timestep so that errors are dominated by spatial discretization. The cost of the continuation procedure itself is also presented, along with its proportion (in percentage) of the overall computational times.}
  \begin{center}
    \begin{tabular}{ccrr|ccrr}
      \multicolumn{4}{c}{AB4} & \multicolumn{4}{c}{RK4}\\\hline
     $N$ & $L^\infty$ error & \multicolumn{1}{c}{Time} & \multicolumn{1}{c}{FC time ($\%$)} &  $N$ & $L^\infty$ error & \multicolumn{1}{c}{Time} & \multicolumn{1}{c}{FC time ($\%$)} \\ \hline
     $20\times20$& $1.55\times10^{-2}$ & 48.48 s & 05.62 s (11.59$\%$) & $20\times20$& $1.55\times10^{-2}$ & 182.32 s & 22.29 s (12.22$\%$) \\
     $40\times40$& $5.02\times10^{-4}$ & 96.60 s & 10.56 s (10.93$\%$) & $40\times40$& $4.99\times10^{-4}$ & 370.72 s & 42.12 s (11.36$\%$) \\
     $80\times80$& $9.53\times10^{-6}$ & 337.62 s & 20.80 s (06.15$\%$)& $80\times80$& $9.33\times10^{-6}$ & 1326.62 s & 83.32 s (06.28$\%$) \\
     $160\times160$& $2.37\times10^{-7}$ & 1035.49 s & 41.92 s (04.05$\%$) &  $160\times160$& $2.38\times10^{-7}$ & 4810.58 s &202.98 s (04.22$\%$) 

    \end{tabular}
    \end{center}

    \end{table*}}
{Indeed, such results imply that the overall FC-based solver is still dominated by the FFT, similar to other spectral solvers, and the cost of the extension procedure is relatively cheap. Such properties, coupled with high-order accuracy, enable fast calculations as can be observed in Table~\ref{tab:table2}, where a comparison is made between the FC-based solver of this work and a second-order central finite difference (FD) scheme for the velocity field and exact solution of Equation~\eqref{eq:MMS} corresponding to $f_v = 4$ Hz, $f_1=f_2 = 1.00$ Hz, $\kappa_1 = -.25$ m/s, $\kappa_2 = .25$ m/s, $D=1\times10^{-2}$ m$^2$/s, and Dirichlet boundary conditions. The solutions are advanced for $200000$ timesteps, and  discretizations for both methods are chosen such that the corresponding solver achieves similar accuracy (two, three, four and five digits). As can be observed, the FC-based solver enables significantly faster computations and significantly less memory requirements than a commonly-used second-order scheme (greater than 17 times faster and greater than 180 times fewer discretization points required in the largest case). Equivalent findings attesting to the advantages of FC-based simulations, in relation to cost or memory demands, have been showcased for various FC-based solvers in prior works comparing against high-order Padé schemes, FD schemes of up to eighth order, discontinuous Galerkin methods, and different orders of finite-volume methods~\cite{albinbruno, amlanipahlevan, ellingthesis,lyonbrunoII,abcc, amlanibruno}.}

  {
    \begin{table*}

      \caption{\label{tab:table2}Comparison of maximum $L^\infty$ errors as well as computational times between FC and a second-order (central) finite difference (FD) method, both evolved using the same time integrator (AB4). The manufactured solution and velocity parameters correspond to $f_v = 4$ Hz, $f_1=f_2 = 1.00$ Hz, $\kappa_1 = -.25$ m/s, $\kappa_2 = .25$ m/s, $D=1\times10^{-2}$ m$^2$/s, and Dirichlet boundary conditions. Errors are computed after $200000$ timesteps by comparison with the exact solution and with a fine-enough timestep so that errors are dominated by spatial discretization.}
      \begin{center}
        \begin{tabular}{ccr|ccr}
          \multicolumn{3}{c}{FC} & \multicolumn{3}{c}{FD}\\\hline
         $N$ & $L^\infty$ error & \multicolumn{1}{c}{Time (s)} & $N$ & $L^\infty$ error & \multicolumn{1}{c}{Time (s)} \\ \hline
         $10\times10$& $1.00\times10^{-2}$ & 41.11 s &   $18\!\times18$& $1.13\times10^{-2}$ & 48.75 s\\
         $18\times18$& $1.03\times10^{-3}$ & 93.94 s &   $55\times55$& $1.15\times10^{-3}$ & 211.68 s\\
         $28\times28$& $1.01\times10^{-4}$ & 164.98 s &   $171\times171$& $1.16\times10^{-4}$ & 462.46 s\\
         $40\times40$& $1.07\times10^{-5}$ & 204.84 s &    $541\times541$& $1.15\times10^{-5}$ & 3645.41 s
        \end{tabular}
        \end{center}
  
        \end{table*}}

\subsection{Assessment of numerical pollution errors: comparison with natural analytical solutions}

{Numerical diffusion (errors in amplitude) and numerical dispersion (phase error, whose source for advection-like problems has been analyzed by~\cite{reviewerb})} of the proposed FC methodology can be studied by considering test problems for (dye) concentrations that travel over arbitrarily long distances  {(note that numerical diffusion refers to solution errors in amplitude, and does not refer to the choice of physical diffusion constant $D$ which determines the exact diffusion properties of the PDE).} For a 2D domain of size $2$ m $\times$ $2$ m with a constant velocity field $\bv(\bx,t) = \bv_0 = (w_1,w_2)^T$ and diffusion constant $D(\bx,t) = D$, the transport of a Gaussian pulse (ball) of unit height centered at $\bx_0 = (0.1,0.1)^T$ is given by
\begin{equation}\label{eq:diffusion1}
  c(\bx,t) = \frac{a}{a+4t} \exp\left[-\frac{(x_1-0.1-w_1t)^2}{D(a + 4t)} -\frac{(x_2-0.1-w_2t)^2}{D(a + 4t)}\right]
\end{equation}
and, for zero diffusion ($D=0$), the solution is given by
\begin{equation}\label{eq:diffusion2}
  c(\bx,t) =  \exp\left[-\frac{(x_1-0.1-w_1t)^2}{b_1} -\frac{(x_2-0.1-w_2t)^2}{b_2}\right],
\end{equation}
where $a,b_1$ and $b_2$ are arbitrary constants.

\begin{figure}
\includegraphics[width=.485\textwidth]{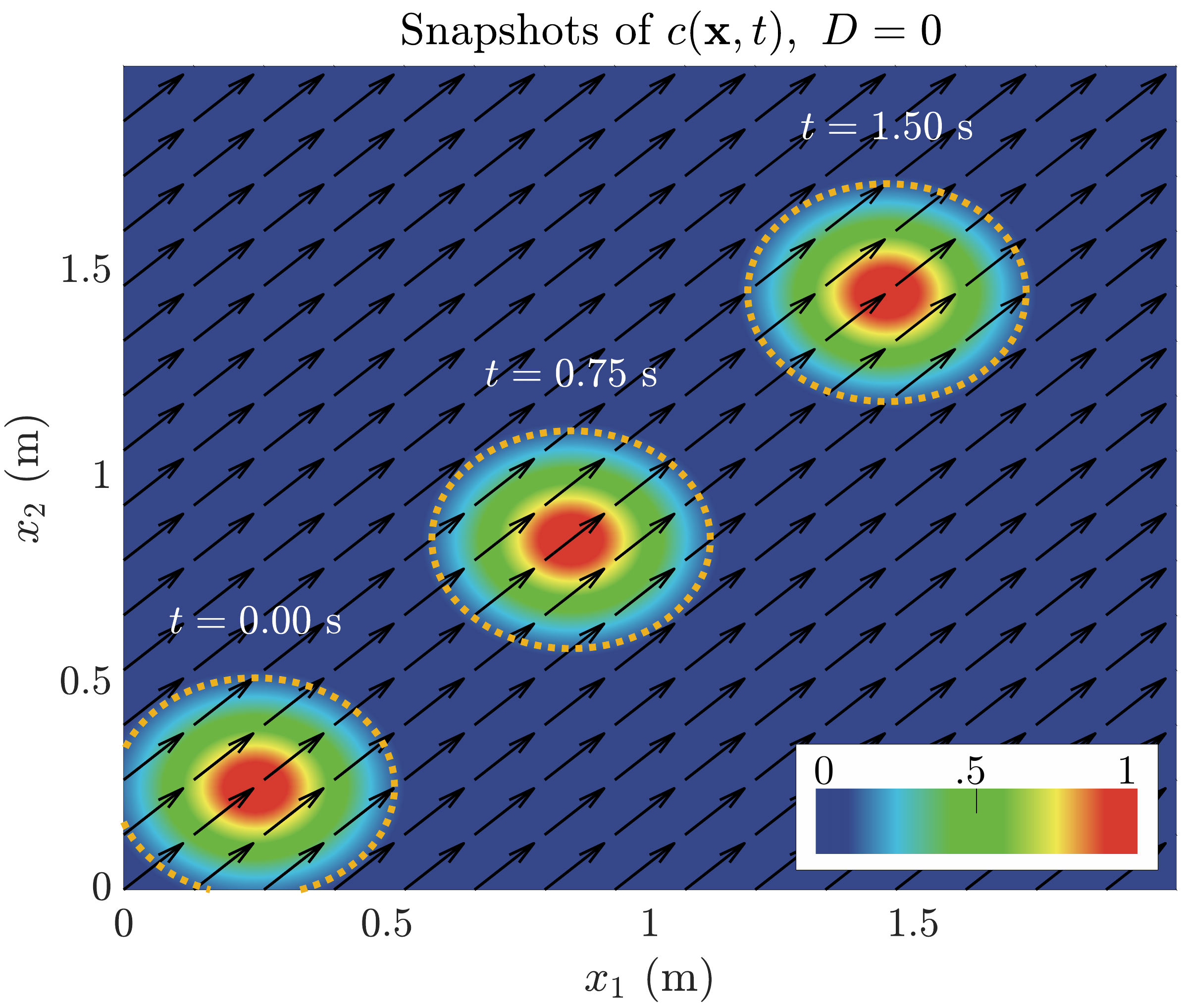}\hfill
\includegraphics[width=.485\textwidth]{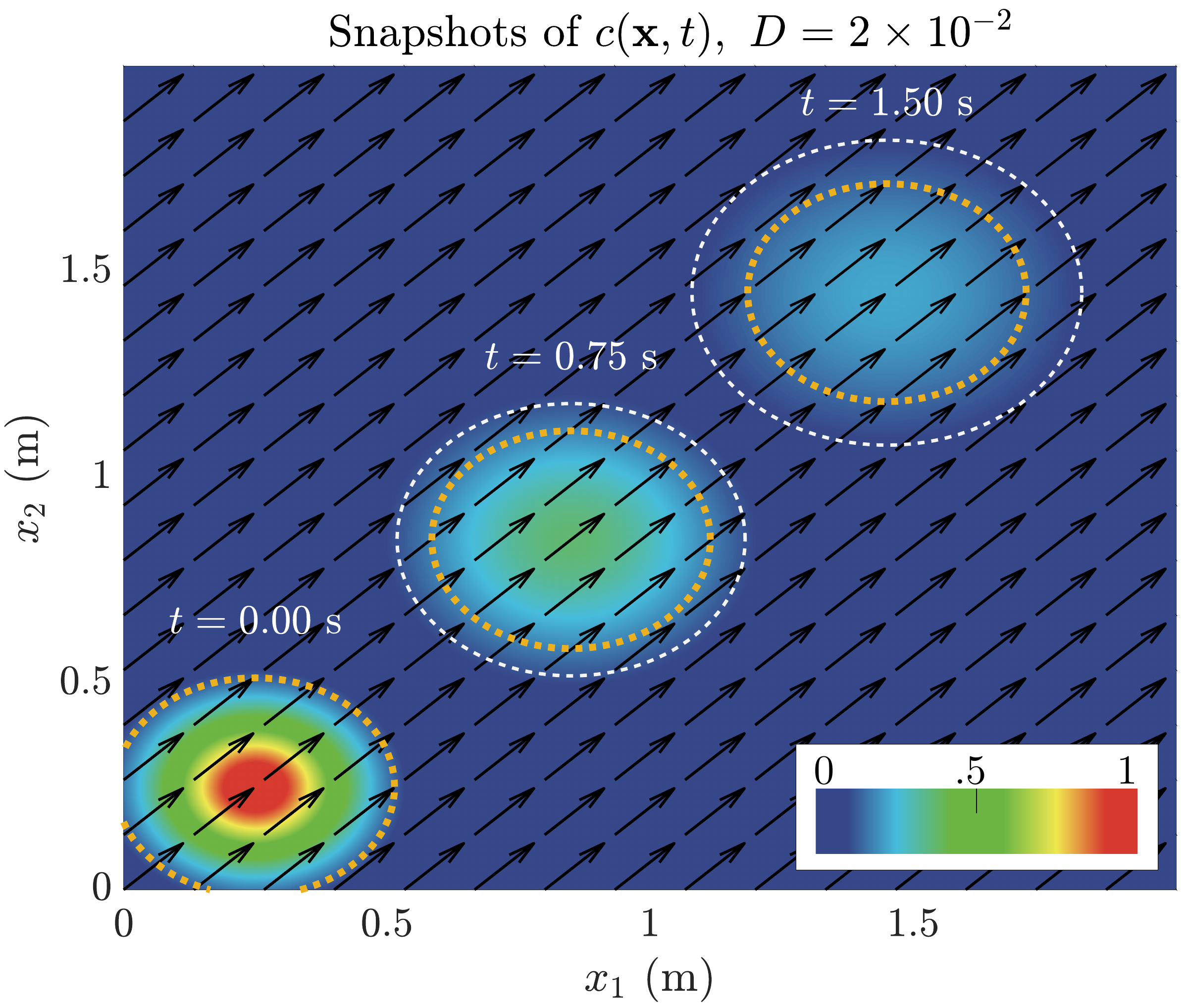}
\caption{\label{fig:fig_4}Numerical snapshots in time using $100\times100$ discretization points to model the evolution of a Gaussian ball of dye for (left) zero physical diffusion  and (right) non-zero physical diffusion  ($D$ is in units of m$^2$/s). }
\end{figure}

\begin{figure}
\includegraphics[width=.485\textwidth]{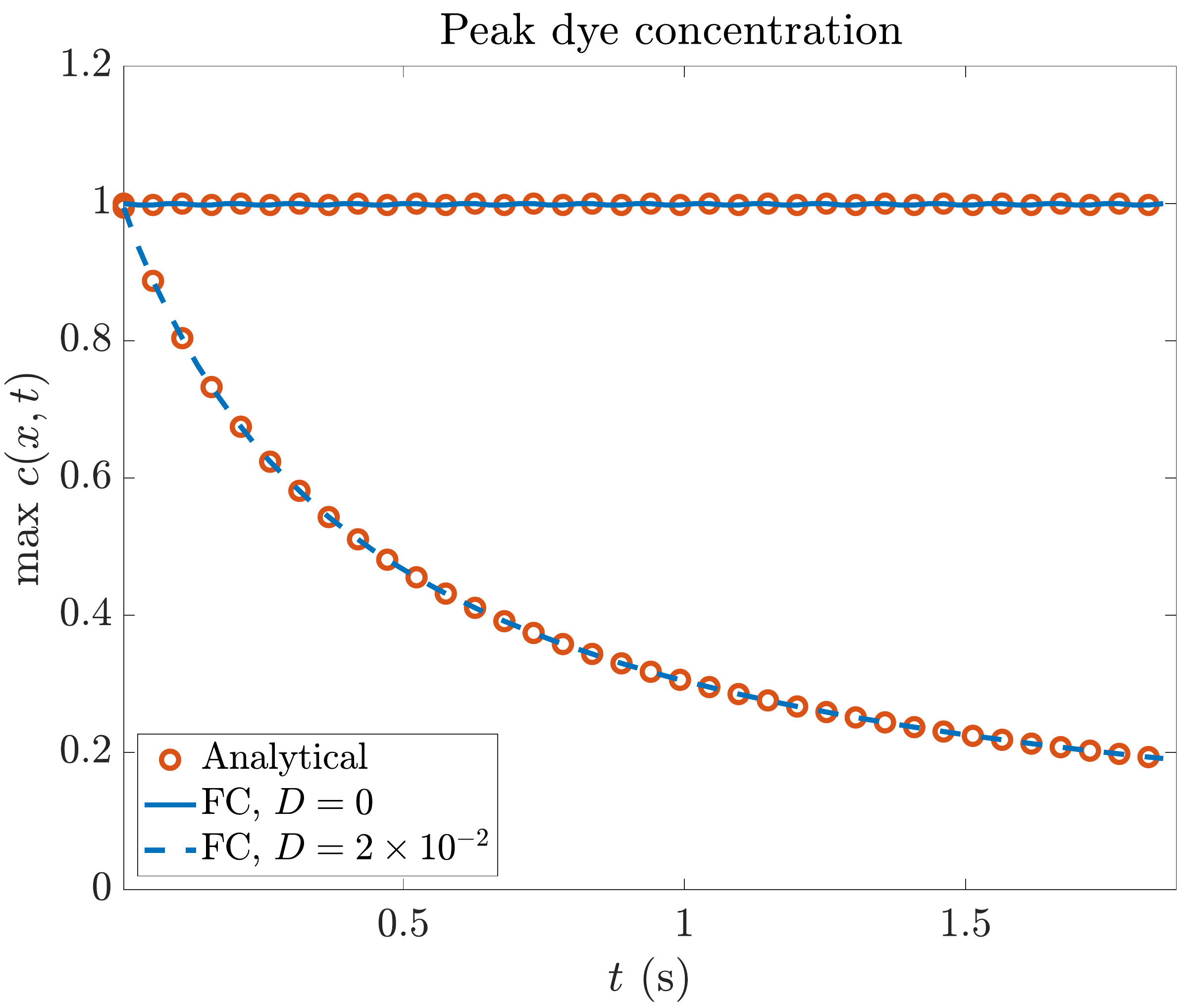}\hfill
\includegraphics[width=.485\textwidth]{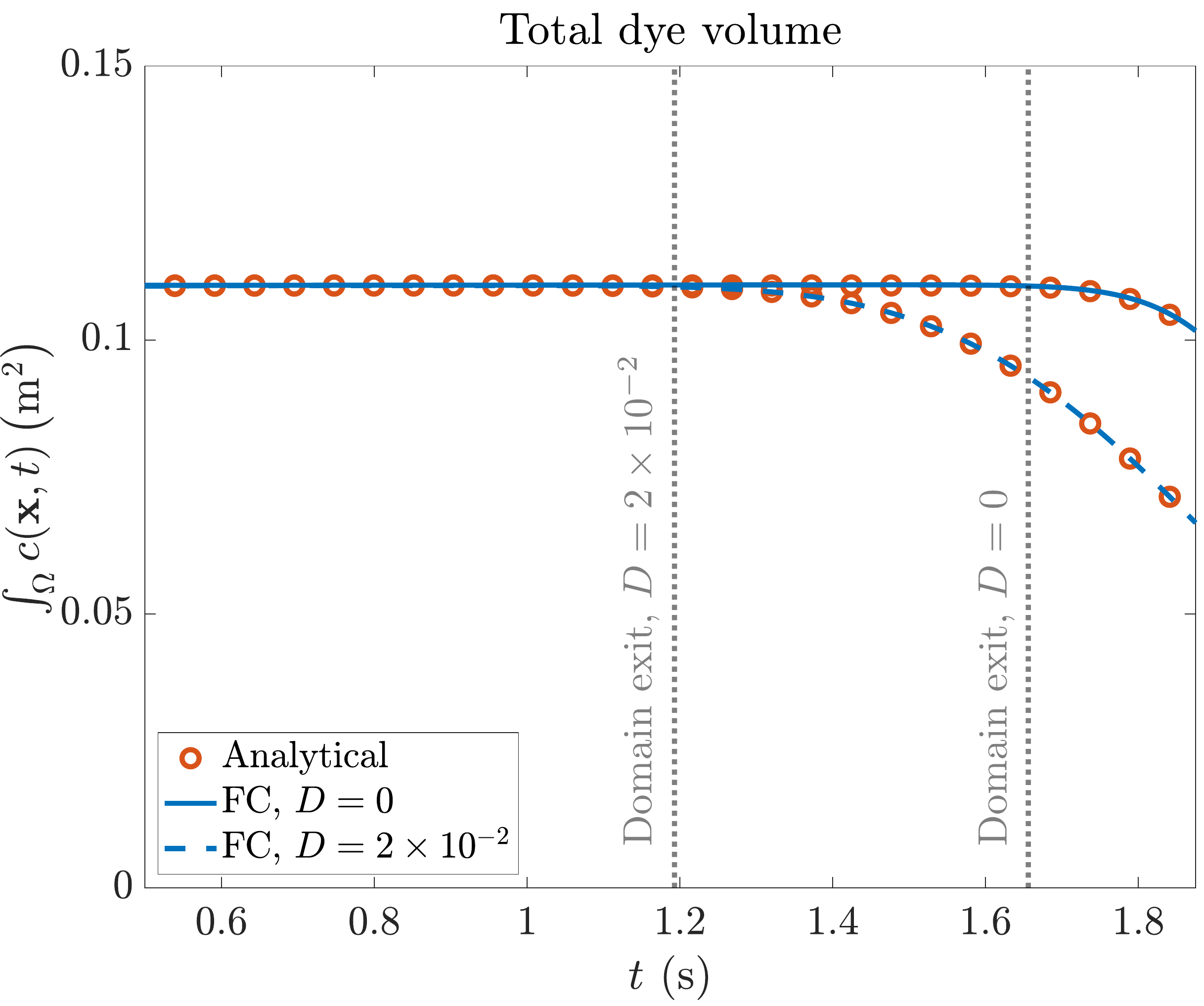}
\caption{\label{fig:fig_5}Peak dye concentration (left) and total dye volume (right) versus time. The dotted lines in the right figure indicate the time at which the dye (non-zero concentration) begins to exit the domain.}
\end{figure}

Together with their initial conditions {(given by evaluating the above at $t=0$)}, it is straightforward to verify that Equations~\eqref{eq:diffusion1} and~\eqref{eq:diffusion2} are analytical (i.e., \emph{not manufactured}) solutions  to the unforced convection-diffusion equation for zero and constant diffusion (respectively) of Equation~\eqref{eq:governing}, i.e.,
$$\displaystyle\pd{c}{t}(\bx,t) + \bv_0\cdot \nabla c(\bx,t) - \nabla\cdot D \nabla c(\bx,t)= 0.$$
Snapshots of the corresponding solutions for $a=1.75$, $w_1=w_2 = 0.8$ m/s, $D$ m$^2$/s and $b_1=b_2 = Da$ are presented in Figure~\ref{fig:fig_4} for zero physical diffusion (left) and non-zero physical diffusion ($D= 2\times 10^{-2}$ m$^2$/s, right). Note that the constants $b_1$ and $b_2$ for the zero-diffusion case are in terms of the diffusive case coefficients in order to affect the same initial Gaussian pulse width for comparative purposes. Figure~\ref{fig:fig_5} presents the corresponding temporal evolution of the peak dye concentration ($\max_{\bx} c(\bx,t)$) as well as total dye volume ($\int_\Omega c(\bx,t) d\bx$) produced by the FC-based solver using a $100\times100$ spatial discretization. The corresponding analytical results marked in the figures indicate excellent {agreement} between analytical and numerical simulations for both the zero physical diffusion and non-zero physical diffusion cases (a weak indication of minimal numerical diffusion). The dotted lines in Figure~\ref{fig:fig_5} (right) demarcate the time at which the dye solution begins to exit the domain, which is expectedly sooner for the physically-diffusive case whose dye concentration has spread wider than in the non-diffusive case.


\begin{figure}
\includegraphics[width=.485\textwidth]{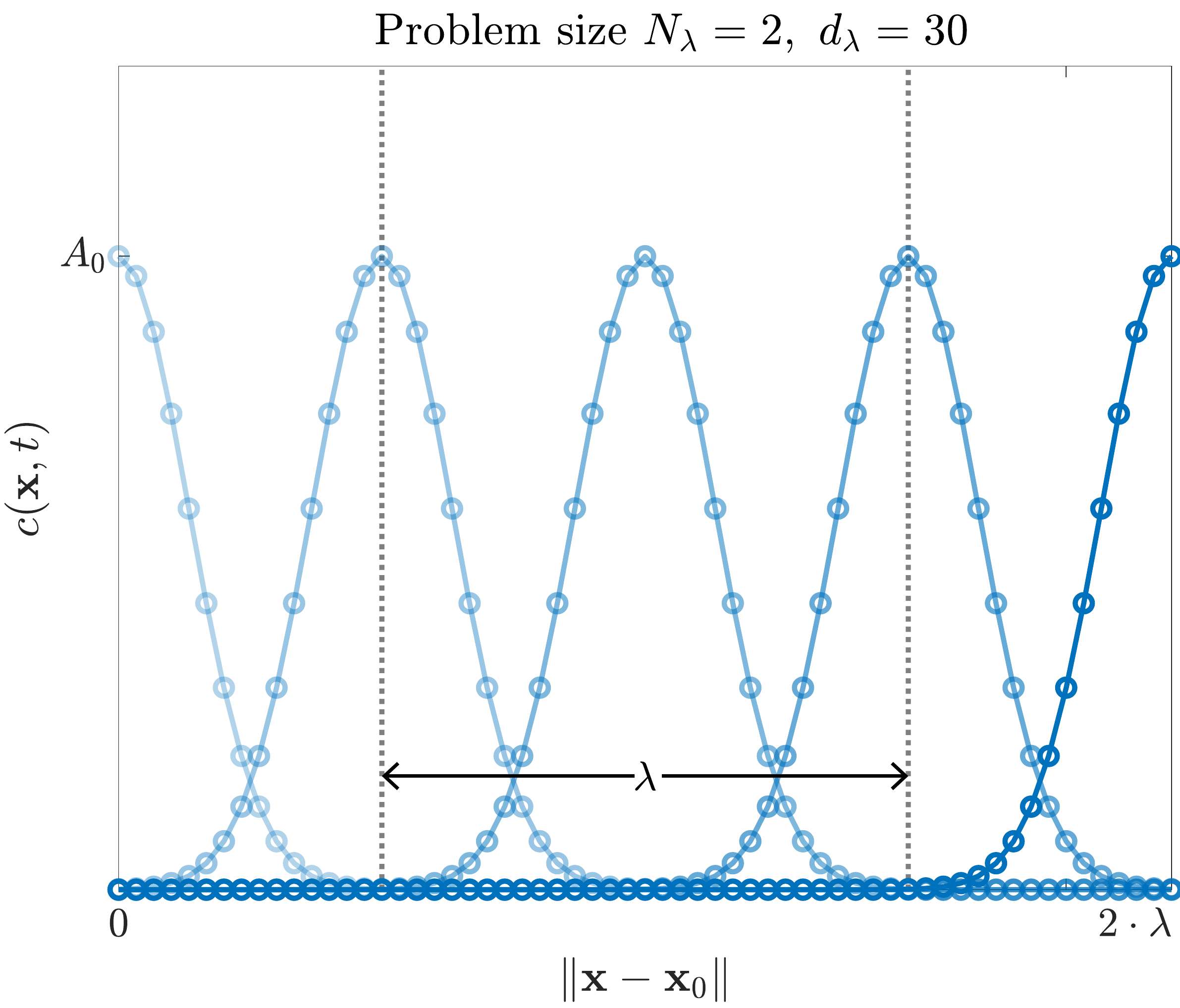}\hfill
\includegraphics[width=.485\textwidth]{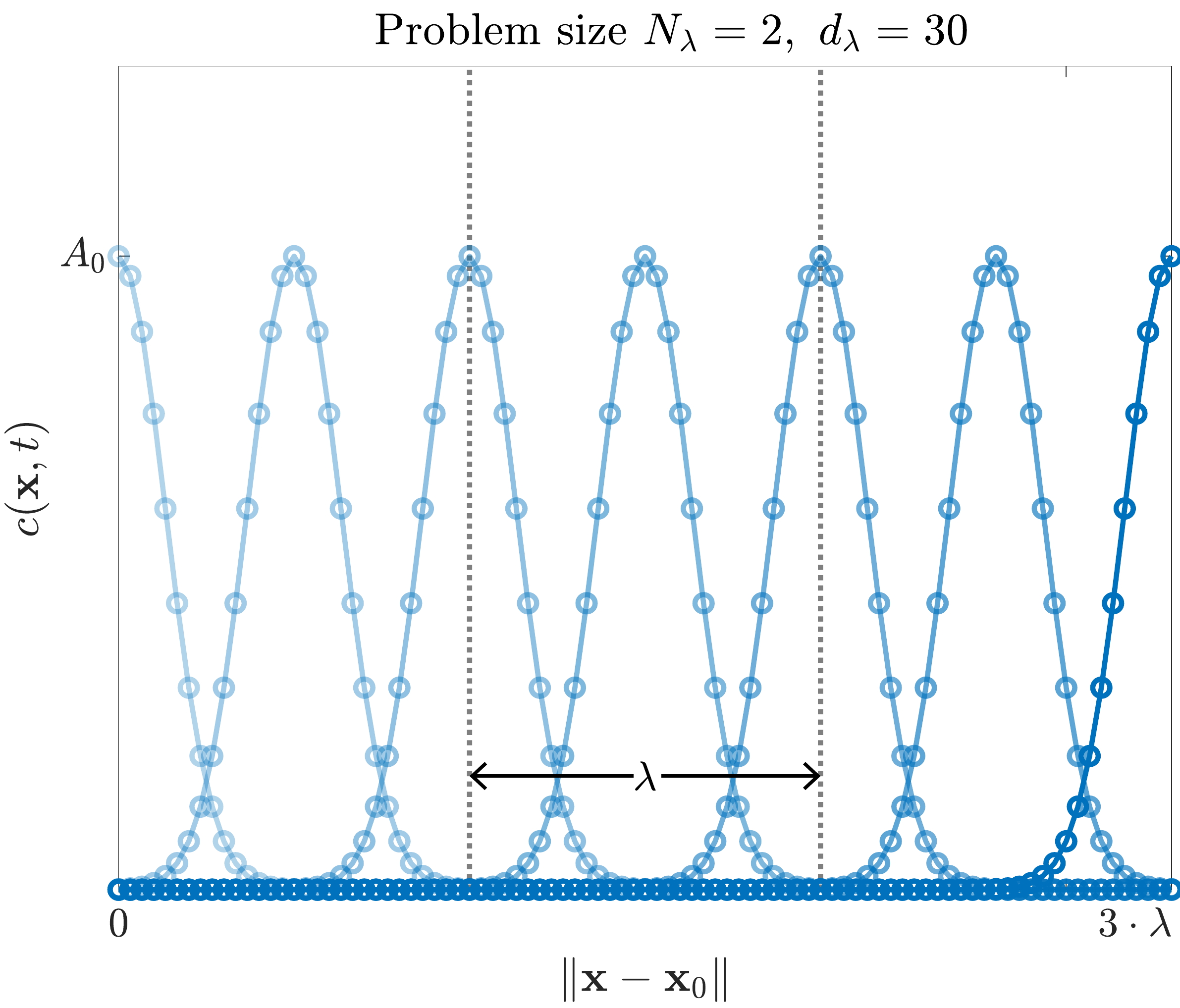}
\caption{\label{fig:fig_6}An illustrative example of the formulation for studying numerical diffusion errors, where a fixed density $d_\lambda$ of points (e.g., $d_\lambda = 30$) is used for resolving an initial dye concentration of characteristic width $\lambda$ (e.g., the effective diameter of the Gaussian-shaped dye ball solution). The problem size $N_\lambda$ corresponds to a systematic increase in distance traversed by the peak dye concentration while maintaining a fixed resolution (i.e., a distance $N_\lambda\cdot\lambda$ discretized by $\Delta x = \lambda/d_\lambda$).}
\end{figure}

Simulation of dye propagation over long distances can be achieved by increasing the distance traveled while maintaining a fixed resolution over a characteristic length $\lambda$. This is {analogous} to wave-based solutions, for which the equivalent problem formulation is achieved by increasing the number of wavelengths across a domain while maintaining a fixed number of points-per-wavelength. Figure~\ref{fig:fig_6} presents a one-dimensional illustration of the problem formulation for studying numerical dispersion errors. Defining $\lambda$ to be the effective initial width (diameter) of the Gaussian ball solution (Equation~\eqref{eq:diffusion2}), and defining $d_\lambda$ to be the number (density) of points discretizing the length $\lambda$, an increase in $N_\lambda$ corresponds to a systematic increase in the distance traveled by the peak dye concentration while maintaining this fixed resolution (i.e., a distance $N_\lambda\cdot\lambda$ discretized by $\Delta x = \lambda/d_\lambda$).

\begin{figure}
\includegraphics[width=.485\textwidth]{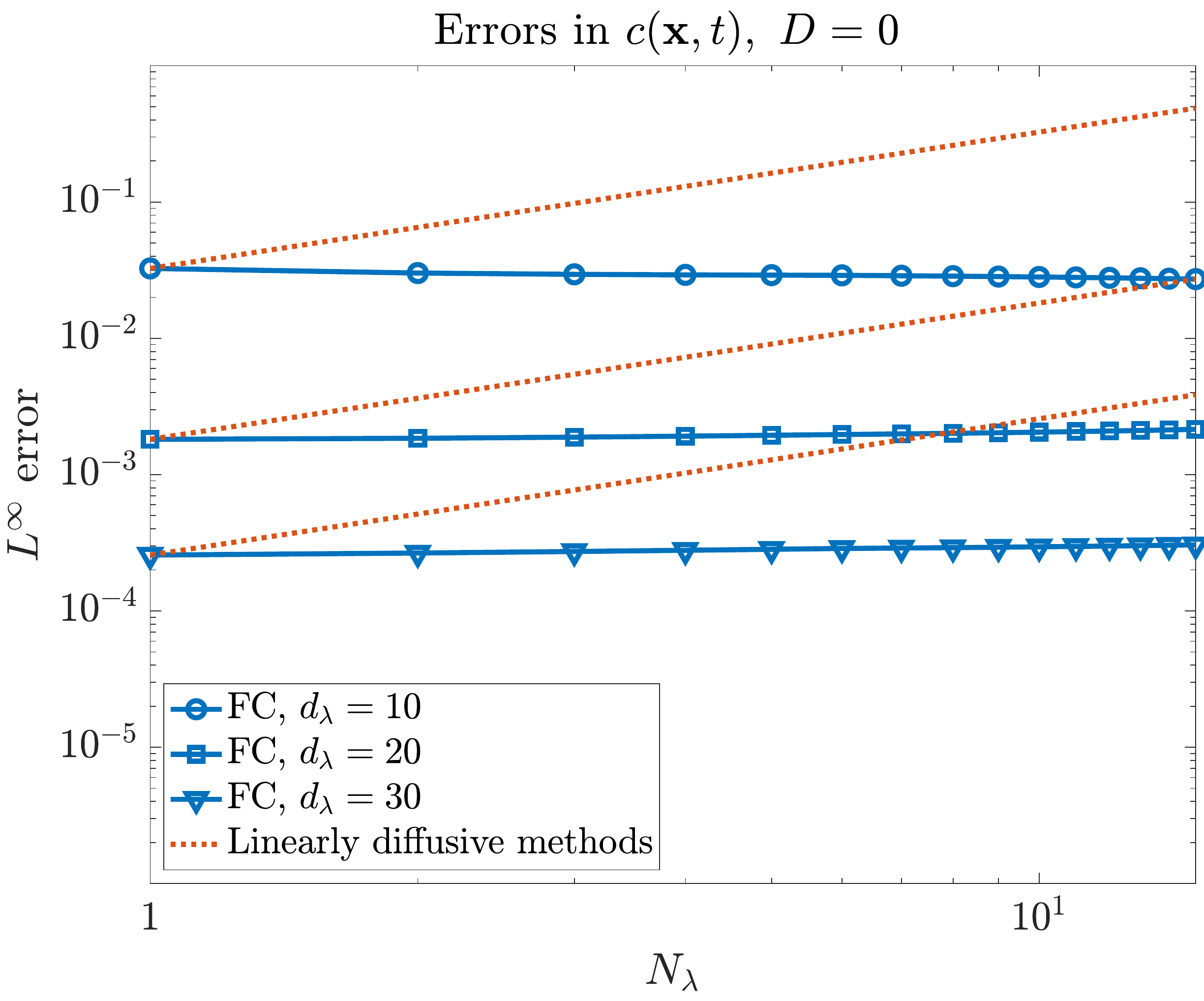}\hfill
\includegraphics[width=.485\textwidth]{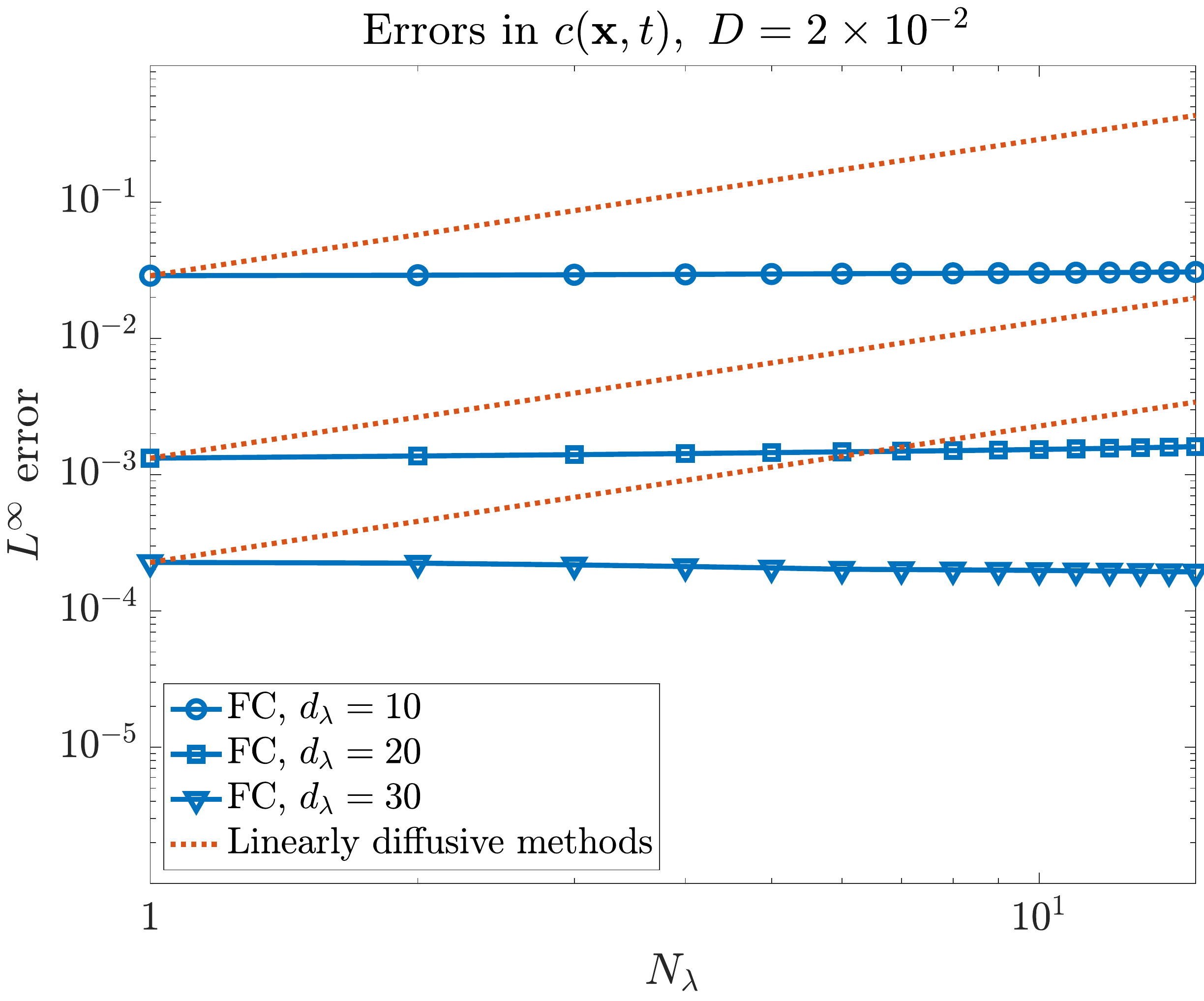}
\caption{\label{fig:fig_7}Maximum $L^\infty$ errors over all space and time for increasing problem size $N_\lambda$ corresponding to (left) zero physical diffusion and (right) non-zero physical diffusion (${D}$ is in units of m$^2$/s). For fixed numbers $d_\lambda =10,20$ or $30$ of points used to discretize the characteristic length $\lambda$, these errors effectively remain constant for arbitrarily long distances traversed by the dye solution. For reference, the red dotted lines represent the expected behaviors of linearly diffusive numerical methods such as those based on low-order finite differences or finite elements~\cite{sankar1998,kalinowska2007,amlanipahlevan,babuska,mullen}.}
\end{figure}

With a timestep chosen small enough so that numerical errors are dominated by those arising from the spatial discretization, Figure~\ref{fig:fig_7} presents maximum
numerical errors over all time and space (as functions of the problem size $N_\lambda$) of the FC-based simulations for both the zero-diffusion and physical-diffusion cases. {The solutions are evolved for the number of timesteps required for the dye ball to pass through the entirety of the domain (which varies based on $N_\lambda$).}
 Each Gaussian ball is discretized by a fixed density of $d_\lambda = 10,~20$ or $30$ points per effective diameter ($\lambda$). The figures are overlaid with the expected behavior of certain numerically diffusive methods~\cite{kalinowska2007,amlanipahlevan,babuska,mullen} such as finite differences (including for convection-diffusion equations~\cite{sankar1998}) or finite elements where, even for small problem sizes $N_\lambda$, larger and larger densities $d_\lambda$ would be necessary in order to produce reasonable accuracies.
 By contrast, the numerical accuracy resulting from the FC-based solutions
 (for both the physically-diffusive and non-diffusive cases) remains essentially constant with increasing $N_\lambda$. The errors are effectively independent of Gaussian ball width (and distance traveled) when the density $d_\lambda$ of spatial points discretizing the width is kept constant---a strong indication that the proposed FC-based methodology incurs effectively no numerical dispersion errors.

\section{\label{sec:casestudies}Physical case studies}

This section {presents} three different physical case studies: 1) a classical 2D example of steady flow over a cavity~\cite{marsden2}; 2) a 3D-axisymmetric example of pulsatile blood flow in a simplified partially-grafted aortic dissection (chosen for its combination of rigid and elastic walls in solid-fluid interactions); and 3) a 3D example of a non-Newtonian fluid in a Fontan graft~\cite{wei2020significance} .

\subsection{\label{sec:cavity}Flow over a cavity (2D)}

\begin{figure}
\includegraphics[width=.485\textwidth]{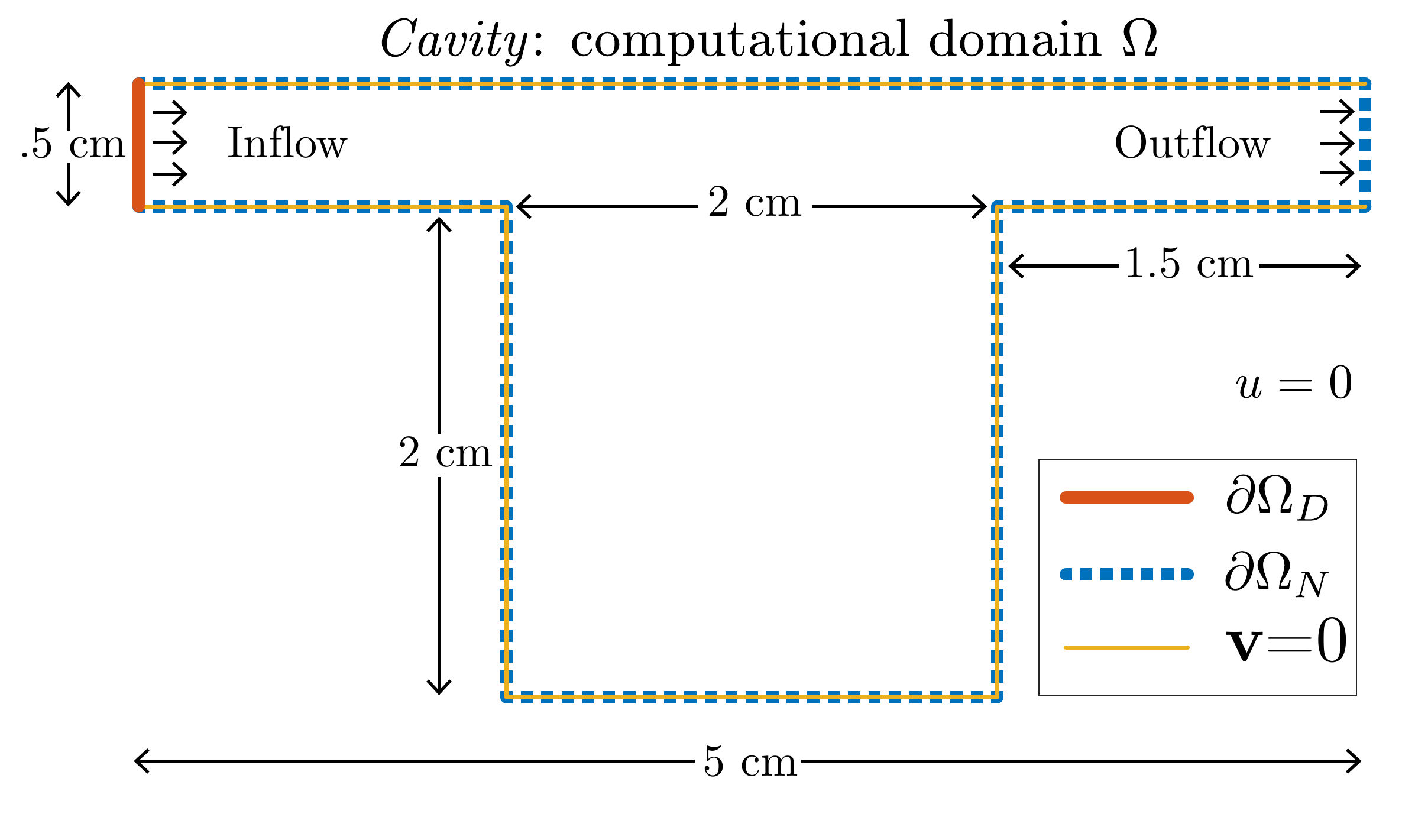}\hfill
\includegraphics[width=.485\textwidth]{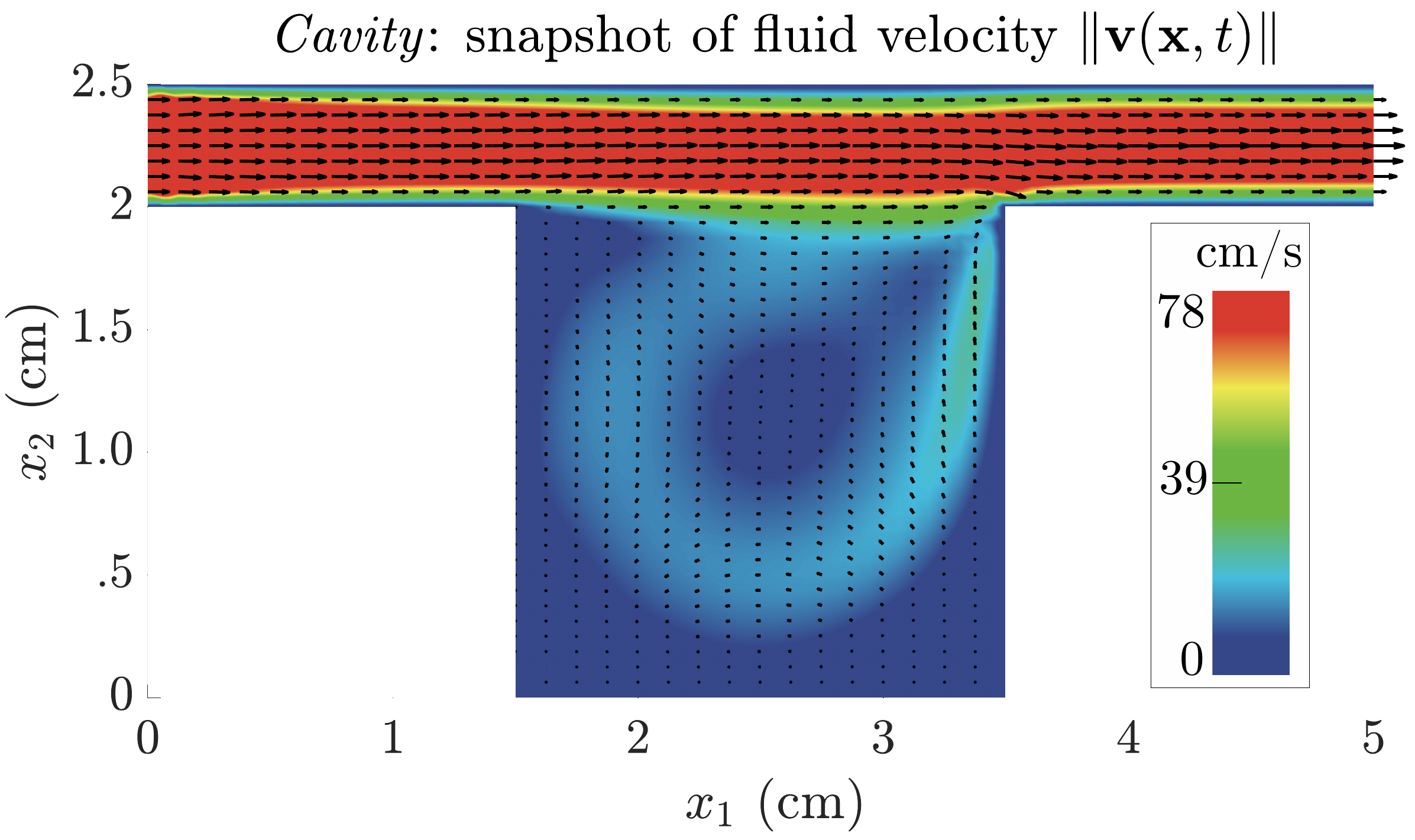}
\caption{\label{fig:fig_8}The computational configuration (left) for the flow over a cavity, where the yellow lines indicate rigid walls where the flow velocity is zero. (Right) A snapshot of the corresponding steady-state velocity field produced by the lattice Boltzmann method.}
\end{figure}

The 2D computational domain $\Omega$ of a cavity underneath a channel cross-flow is given in Figure~\ref{fig:fig_8} (left) and corresponds to an example previously proposed for residence time calculations~\cite{marsden2}. The fluid velocities are provided by the lattice Boltzmann solver described earlier employing fluid density and viscosity as $1.18\times10^{-3}$ g/cm$^3$ and $1.82\times10^{-4}$ g/(s$\cdot$cm), respectively, with a constant uniform velocity of $77.1$ cm/s imposed at the inlet and zero traction at the outlet (corresponding to $\mathrm{Re}=1000$). A snapshot of the resulting steady-state solution in the domain is given in Figure~\ref{fig:fig_8} (right), where flow appears to strongly recirculate in the cavity (while remaining stagnant in the bottom cavity corners).

\begin{figure}
\includegraphics[width=.485\textwidth]{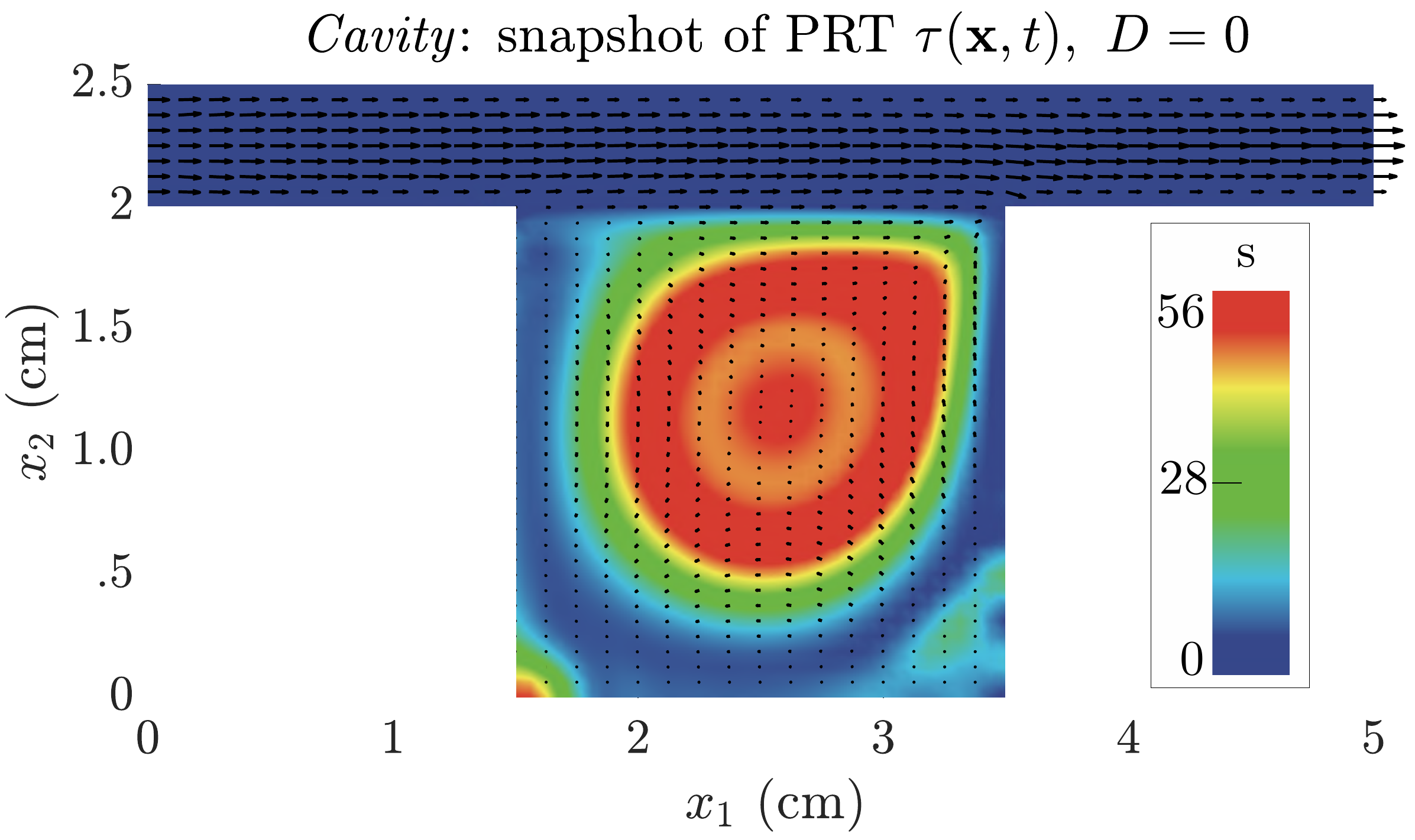}\hfill
\includegraphics[width=.485\textwidth]{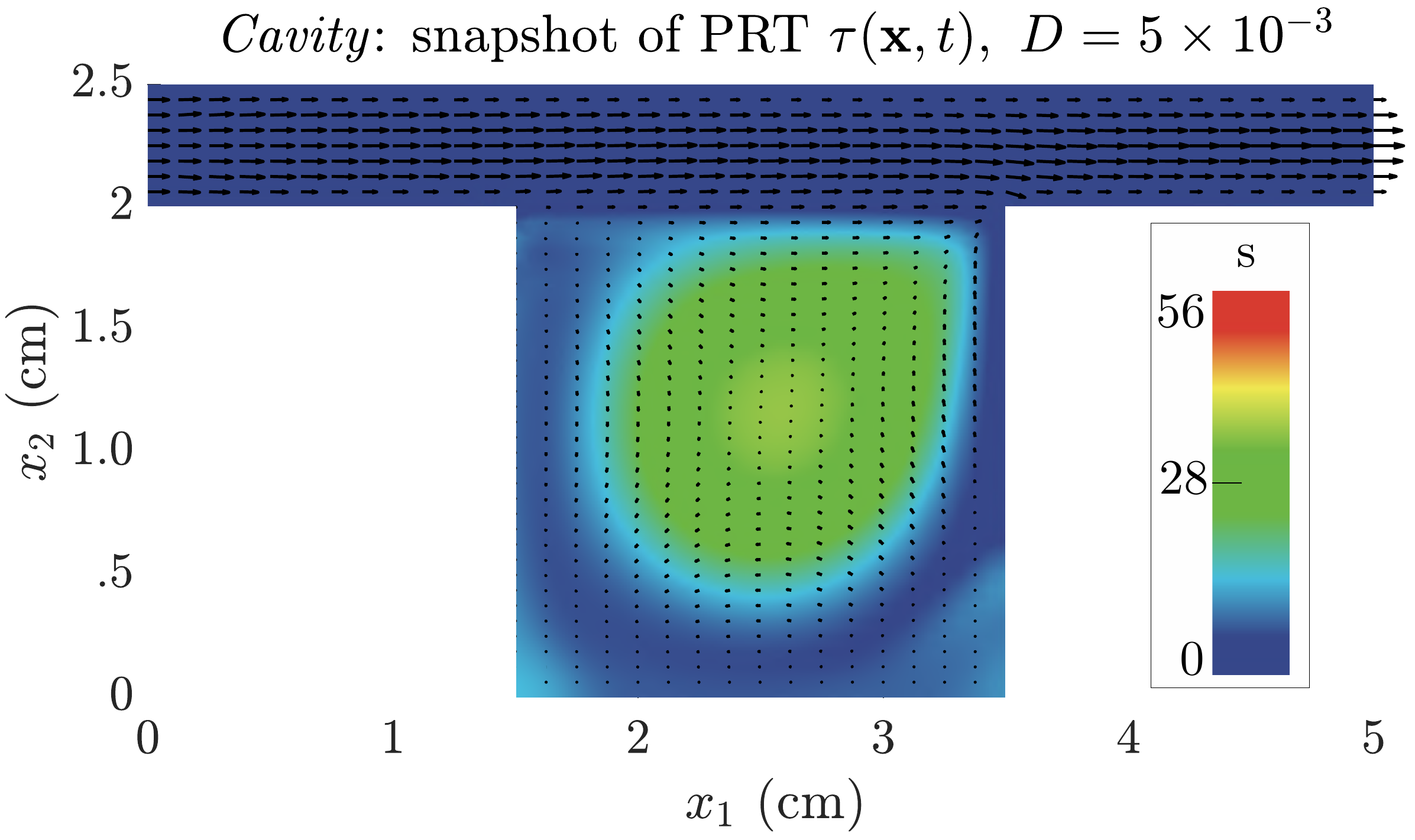}
\caption{\label{fig:fig_9}Numerical snapshots of particle residence times for (left) zero physical diffusion and (right) non-zero physical diffusion ($D$ is in units of cm$^2$/s).}
\end{figure}

Using a spatial discretization of $80\times40$ points (corresponding to $\Delta x_1 = \Delta x_2 = 1/16$ cm) and a timestep of $\Delta t = 0.04$ ms for the convection-diffusion PDE, the FC-based particle residence time calculations for cases with zero physical diffusion and non-zero physical diffusion ($D=D_{x_1}=D_{x_2} = 5\times10^{-3}$ cm$^2$/s) are presented in Figure~\ref{fig:fig_9}. The simulation results demonstrate that the fluid gets trapped in the cavity and the stagnation corners as expected. In order to further validate the results of FC-based simulations with those provided by others~\cite{marsden2}, Table~\ref{tab:table3} additionally presents various measures/metrics computed over the entire domain as well as just the cavity. Defining $T$ as the effective length of time of a simulated ``cycle" (of which there are $N_\text{cycles}$) and $V_{\Omega_i} =\int_{\Omega_i}d\Omega_i$ as the volume (area) of the region of interest $\Omega_i$ (the full domain or only the cavity), two additional residence time measures $RT_1$ and $RT_2$ have been proposed~\cite{marsden2} as
\begin{equation}\label{eq:RT}
RT_1 = \frac{1}{T}\int_{(N_\text{cycles}-1)T}^{N_\text{cycles}T} \frac{1}{V_{\Omega_i}} \int_{\Omega_i} \tau(\bx,t) d\Omega dt \quad \text{and} \quad RT_2 = \displaystyle\frac{V_{\Omega_i}}{\frac{1}{T} \displaystyle\int_{(N_\text{cycles}-1)T}^{N_\text{cycles} T} \int_{\Gamma_i} \bv(\bx,t)\cdot \bn d\Gamma dt},
\end{equation}
where the denominator in $RT_2$ represents the average flow \emph{into} the region of interest (e.g., across the boundary $\Gamma_i$ of the inlet or cavity opening where $\bn$ is the outward unit normal). {Here, $RT_1$ measures residence time as a space-time average over the last cycle; and $RT_2$ is a residence time measure based only on the particles that are entering and leaving the region of interest~\cite{marsden2}.} Table~\ref{tab:table3} presents the corresponding values of these measures produced by FC-based simulations, which are well in agreement with those provided by others~\cite{marsden2}. The significance of these metrics has been discussed elsewhere~\cite{marsden2}.

\begin{table*}
\caption{\label{tab:table3}Various residence time measures {in regions $\Omega_i$ of interest for flow over a cavity, where: $V$ is the total volume of the corresponding region $\Omega_i$; $\int_{\Omega_i} \tau d\Omega_i$ is a ``total volume" of residence times; $RT_1$ is the space-time average of the residence time over the last cycle; and $RT_2$ is a measure based only on the particles that are entering and leaving the region of interest (the latter two are calculated by Equation~\eqref{eq:RT}).}}
\begin{center}
\begin{tabular}{ccccccc}
 $\Omega_i$&$V_{\Omega_i}$ (cm$^2$)&$\int_{\Omega_i} \tau d\Omega_i$ (cm$^2$ s)
&$RT_1$ (s) & $RT_2$ (s) & $RT_2/RT_1$\\ \hline
 Full & $6.5$ & 135.9 & 20.88 & 0.17 & 0.0081\\
Cavity & $4.0$ & 134.5 & 33.57 & 6.28 & 0.1870\\
\end{tabular}
\end{center}
\end{table*}

\begin{figure}
\includegraphics[width=.485\textwidth]{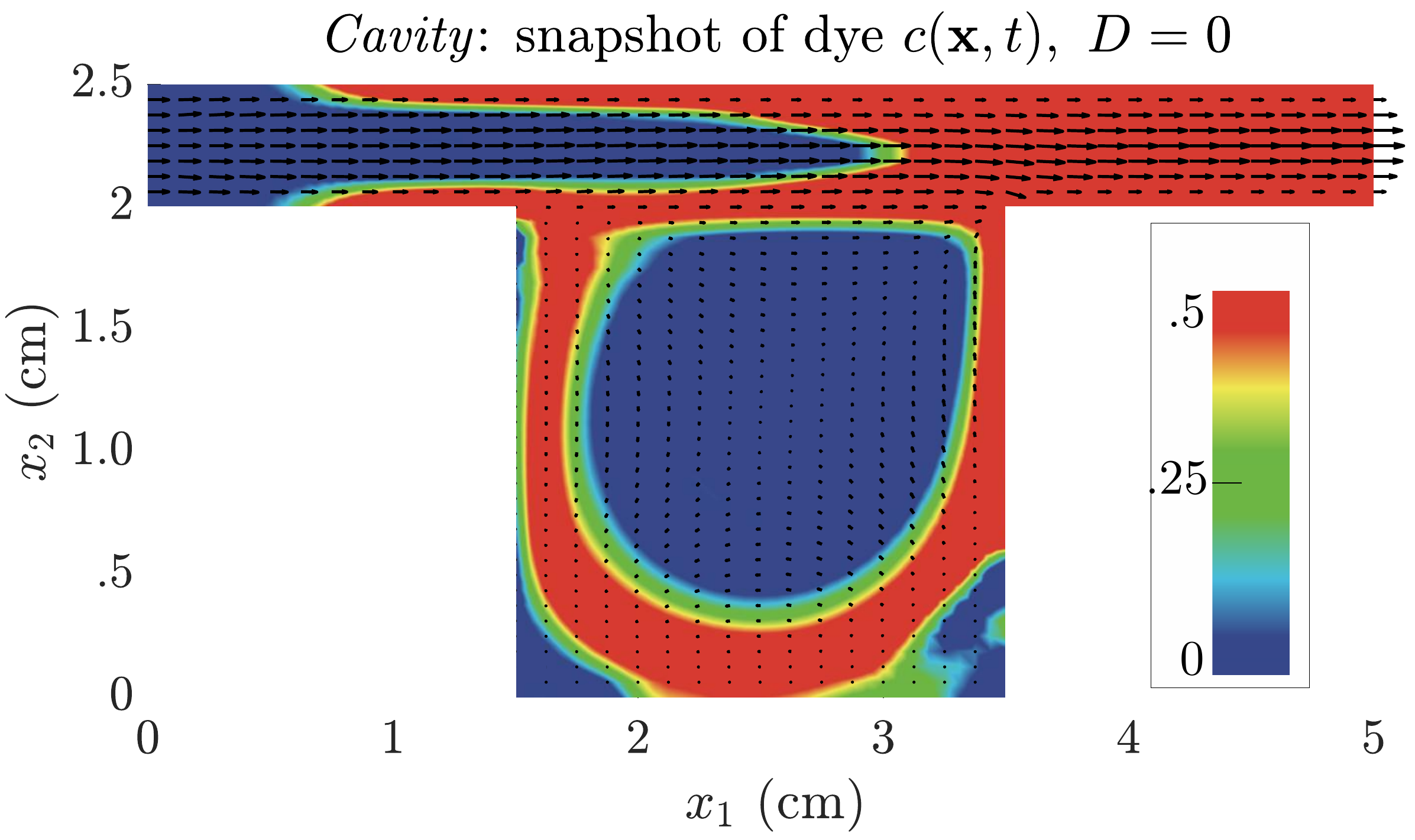}\hfill
\includegraphics[width=.485\textwidth]{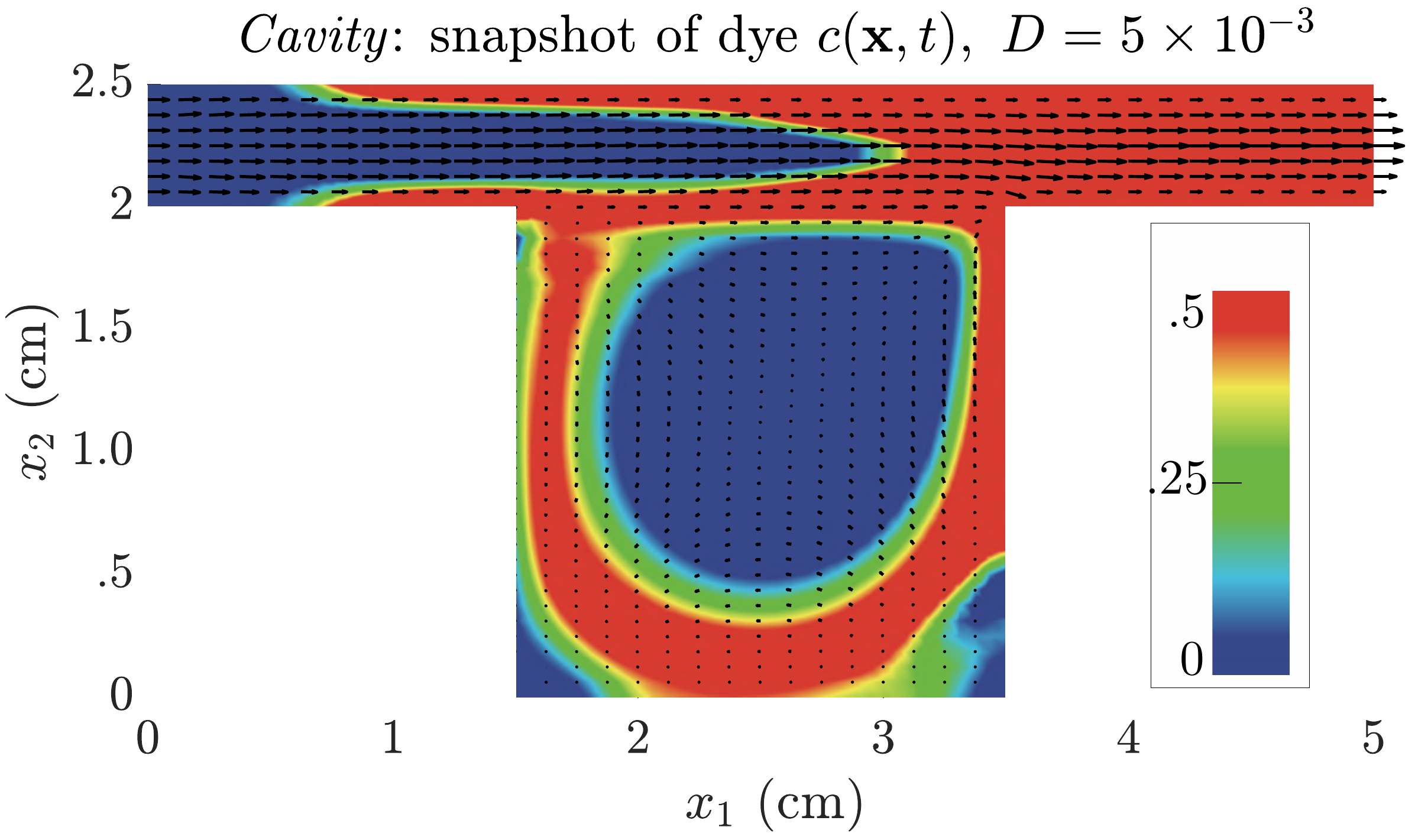}
\caption{\label{fig:fig_10}Numerical snapshots of the concentration at $t=0.03$ s after a dye injection ceases for (left) zero physical diffusion and (right) non-zero physical diffusion ($D$ is in units of cm$^2$/s).}
\end{figure}

\begin{figure}
\includegraphics[width=.485\textwidth]{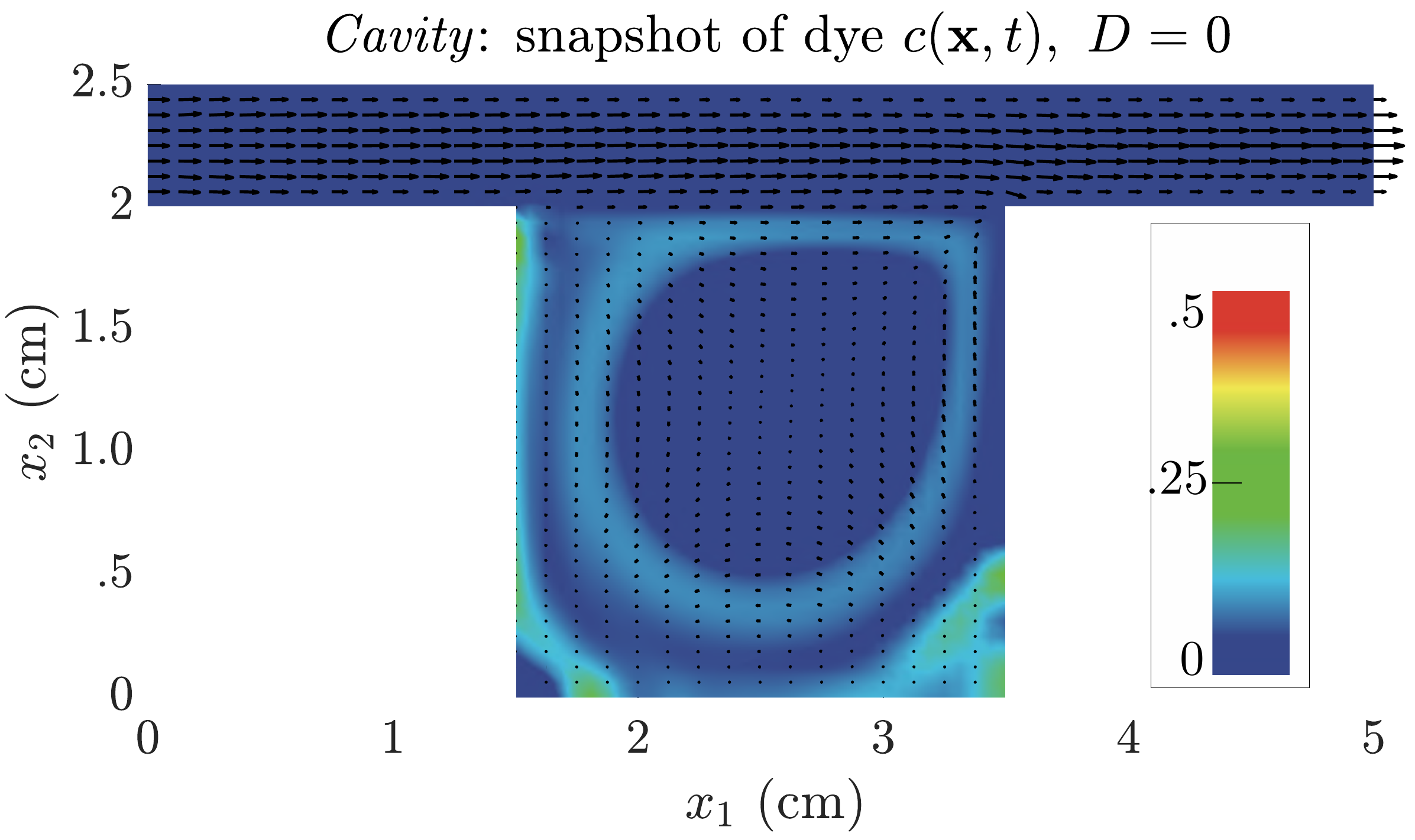}\hfill
\includegraphics[width=.485\textwidth]{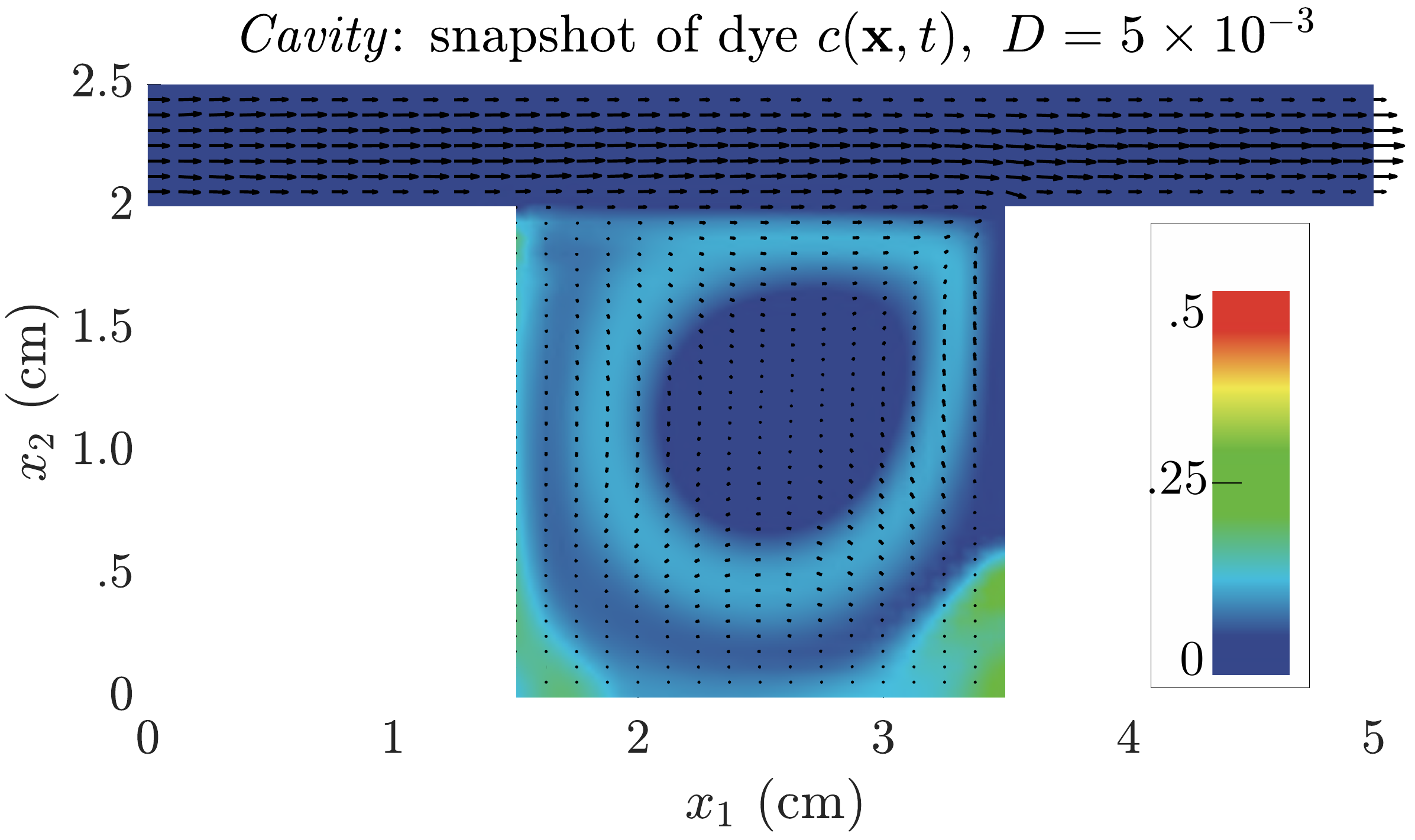}
\caption{\label{fig:fig_11}Numerical snapshots of the concentration at $t=22.32$ s after a dye injection ceases for (left) zero physical diffusion and (right) non-zero physical diffusion ($D$ is in units of cm$^2$/s).}
\end{figure}

Figure~\ref{fig:fig_10} presents snapshots at $t=2.03$ s of an FC-based simulation of a dye injected across the inlet for $2$ s (or 2 cycles) for both zero physical diffusion and non-zero physical diffusion ($D=D_{x_1}=D_{x_2} = 5\times10^{-3}$ cm$^2$/s), where the latter illustrates early signs of the physical diffusion. Figure~\ref{fig:fig_11} presents corresponding snapshots at $t=24.32$ s, where the physically-diffusive configuration leads to larger concentrations of dye particles being trapped in the cavity (particularly the bottom-left stagnation corner, where a non-diffusive dye may take much longer to reach the area---if it does all).

\subsection{\label{sec:dissection}Pulsatile flow through an aortic dissection (3D axisymmetric)}

\begin{figure}
  \begin{center}
\includegraphics[width=.97\textwidth]{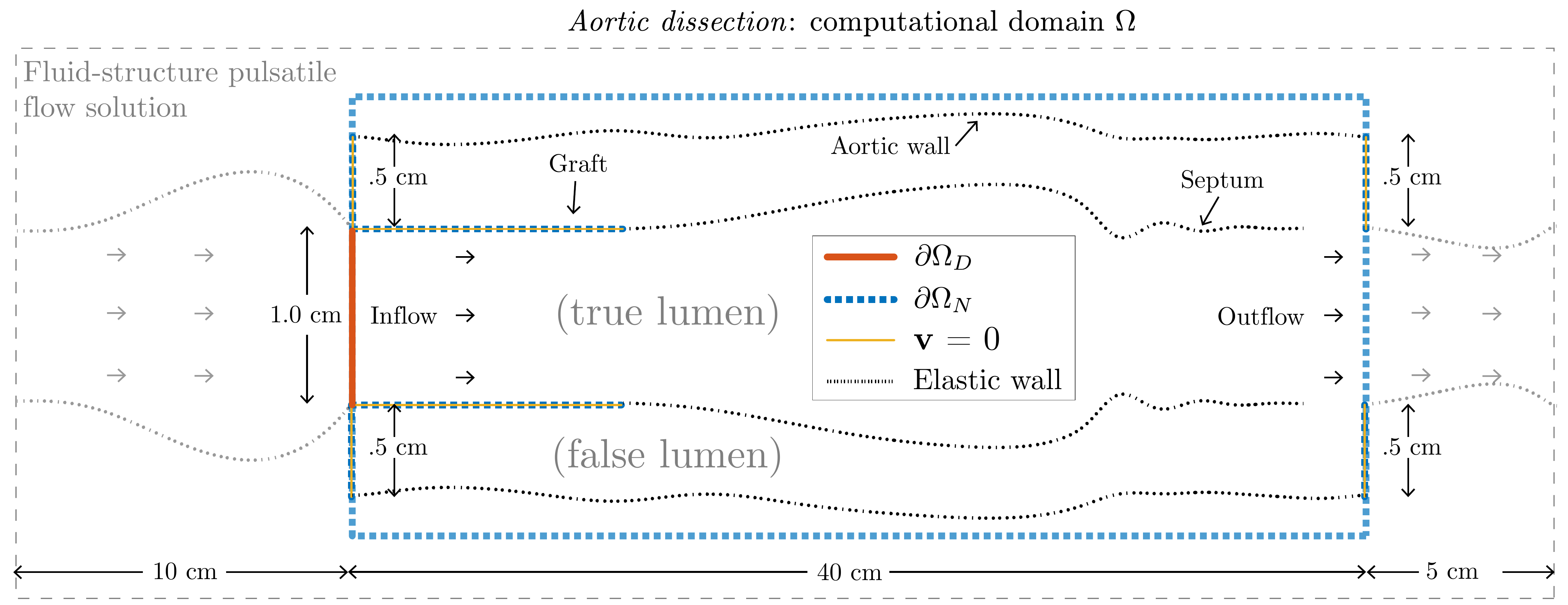}
\end{center}
\caption{\label{fig:fig_12}The computational configuration for the partially-grafted aortic dissection, where the yellow lines indicate rigid walls for which flow velocity is zero, and white lines indicate the elastic boundaries of the dissection and outer aorta.}
\end{figure}

\begin{figure}
  \begin{center}
\includegraphics[width=.72228\textwidth]{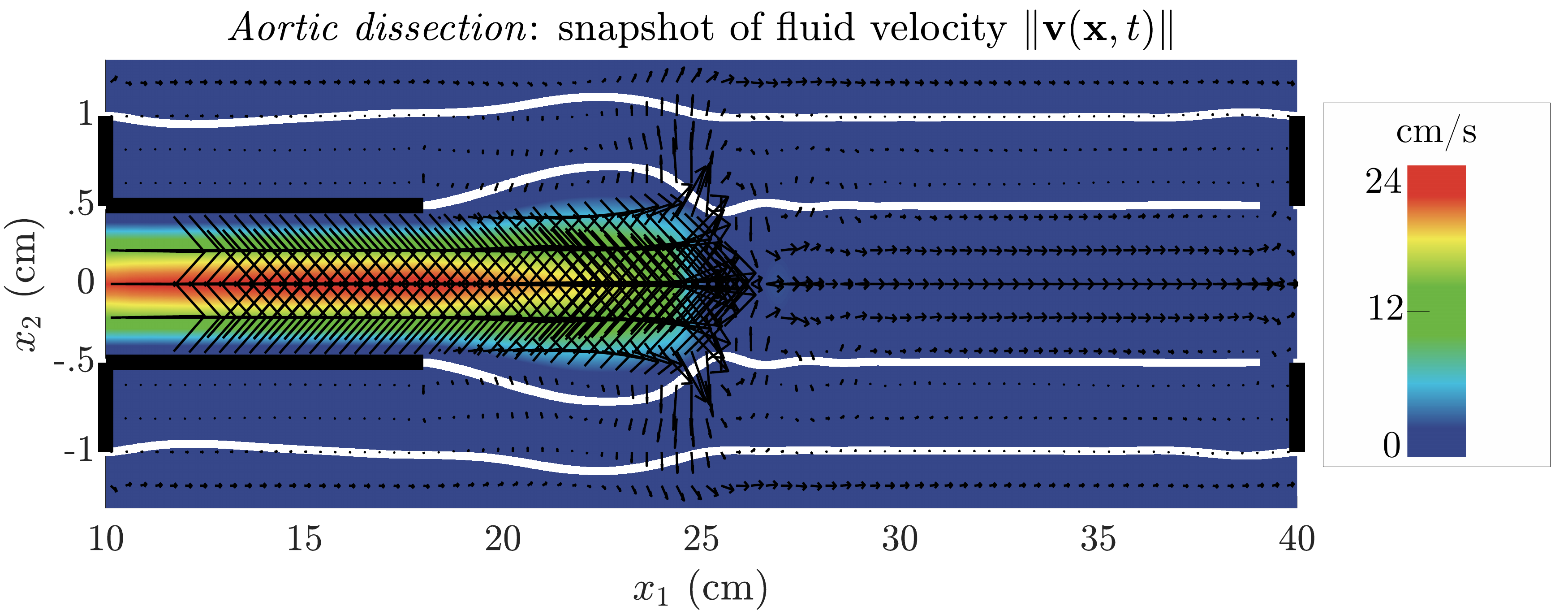}\\[.5em]
\includegraphics[width=.72228\textwidth]{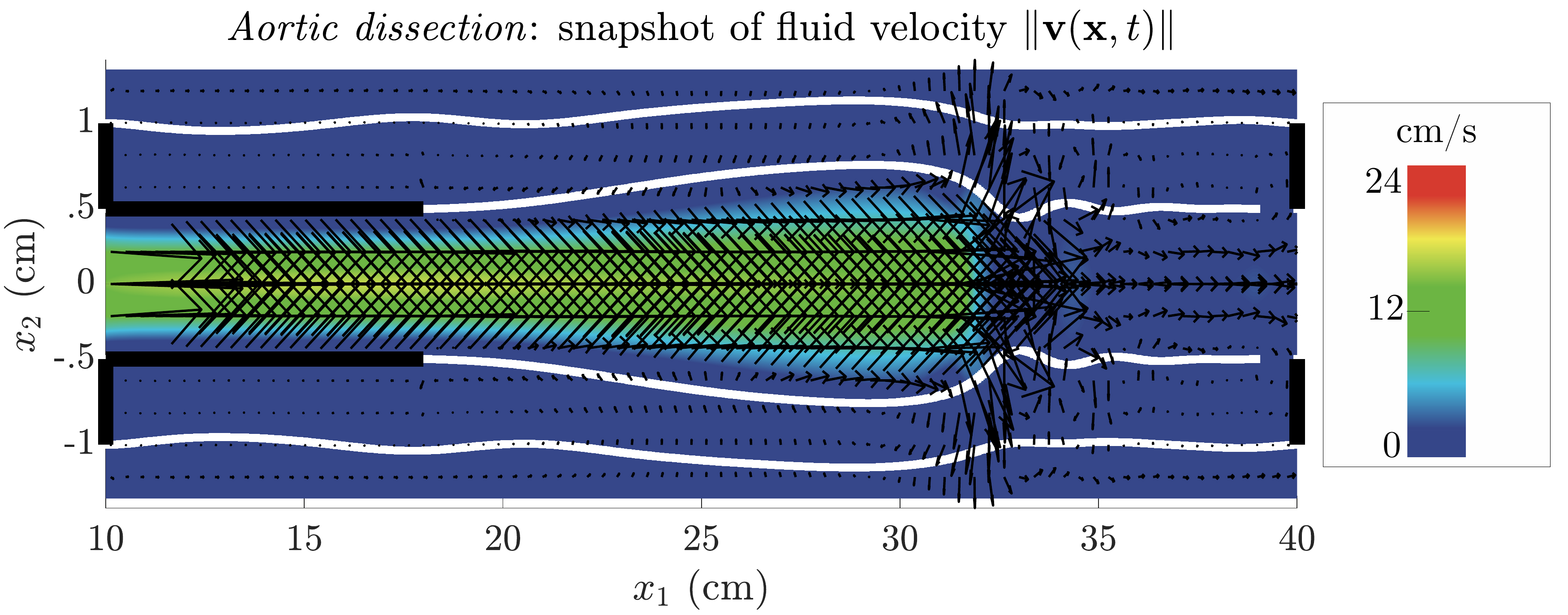}\\[.5em]
\includegraphics[width=.72228\textwidth]{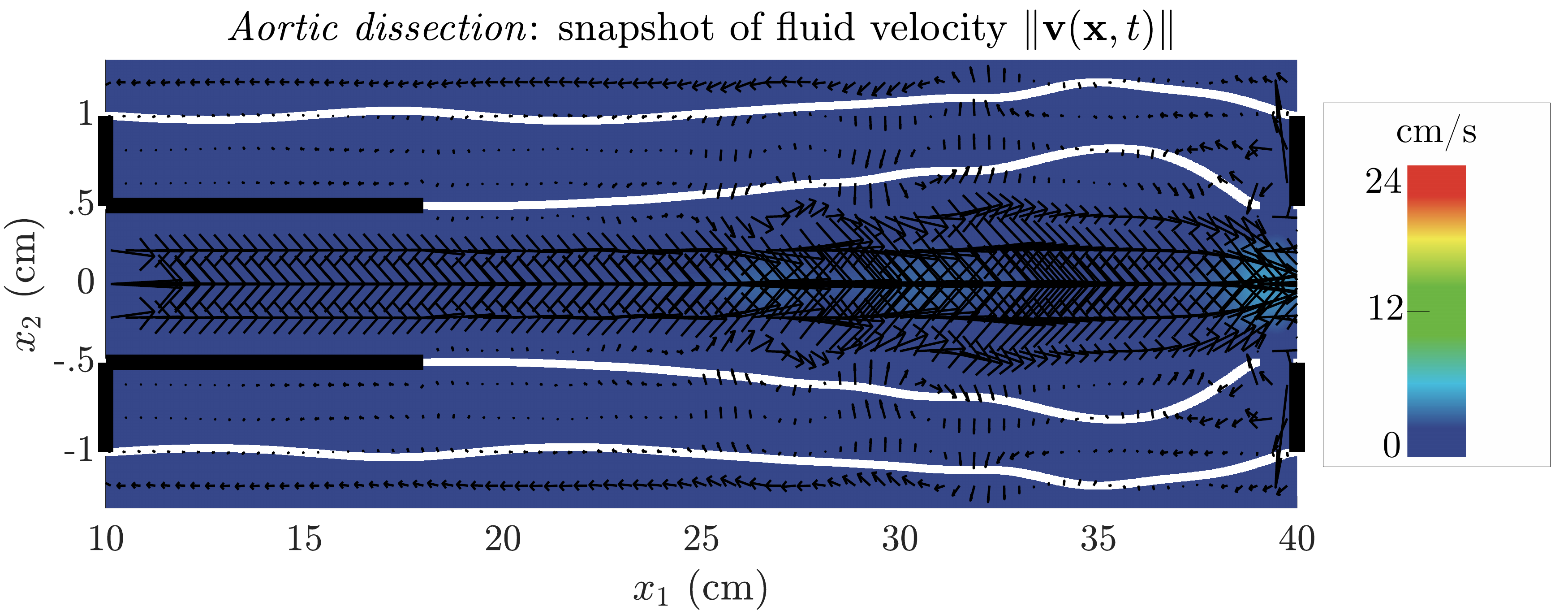}
\end{center}
\caption{\label{fig:fig_13}Numerical snapshots of the (pulsatile) flow velocity at various times during a single cardiac cycle (1 s) produced by an FSI immersed boundary-lattice Boltzmann method. Black lines indicate rigid walls, and white lines indicate compliant (elastic) walls.}
\end{figure}

Calculation of particle residence times in a false lumen of an aortic dissection is physiologically significant since promoting thrombus (blood clot) formation in a false lumen (linked to PRTs) is of clinical interest~\cite{erbel1993effect}. The 3D axisymmetric computational domain $\Omega$ of a simplified partially-grafted aortic dissection is illustrated in Figure~\ref{fig:fig_12}, where both rigid walls and compliant walls (governed by the elastic PDEs of Equation~\eqref{eqn:structure}) are indicated. {The outer aortic wall is submerged in a surrounding body fluid in order to accomodate the immersed boundary algorithm for the the lattice Boltzmann-based fluid-structure solver (Section \ref{LBMsec}),} {which} is employed {here} to simulate a physiological pulsatile flow and generate a flat velocity profile at the aortic inlet corresponding to a heart rate of $60$ bpm and a cardiac output of $2.3$ L/min. Fluid parameters correspond to $\mathrm{Wo}=12$ and $\mathrm{Re}=900$. Snapshots at various times of the pulsatile flow in the domain during a single cardiac cycle (of length $1$ s) is presented in Figure~\ref{fig:fig_13}.

\begin{figure}
  \begin{center}
\includegraphics[width=.72228\textwidth]{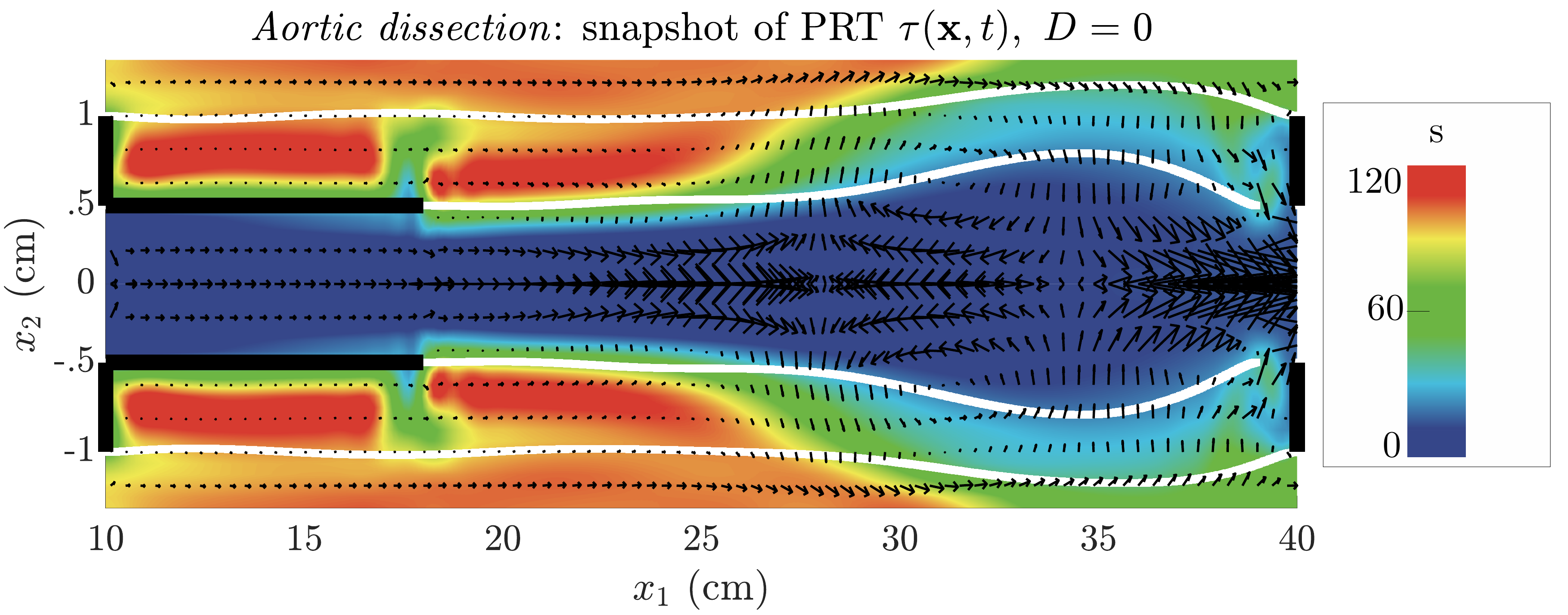}
\end{center}
\caption{\label{fig:fig_14}A numerical snapshot of particle residence times for zero physical diffusion (after many dozens of cardiac cycles). Black lines indicate rigid walls, and white lines indicate compliant (elastic) walls.}
\end{figure}

Using a spatial discretization of $480\times22$ points (corresponding to $\Delta x_1 = \Delta x_2 = 1/16$ cm) and a timestep of $\Delta t = 0.1$ ms for the convection-diffusion PDE, the FC-based particle residence time calculation after many dozens of cardiac cycles {(to ensure the system reaches steady-state oscillatory conditions)} for zero physical diffusion is presented in Figure~\ref{fig:fig_14}. As expected, there are high residence times within the false lumen (between the septum and the aortic wall), particularly within the rigid grafted area.

\begin{figure}
  \begin{center}
\includegraphics[width=.72228\textwidth]{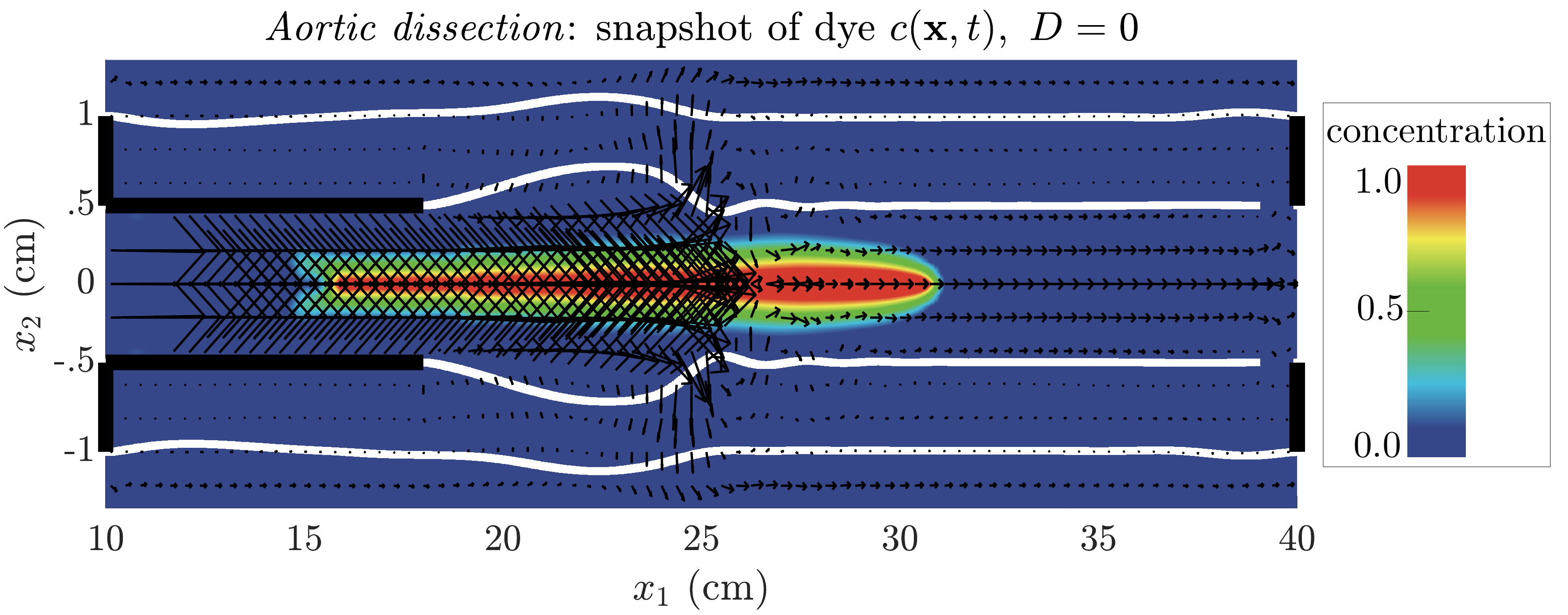}\\[1em]
\includegraphics[width=.72228\textwidth]{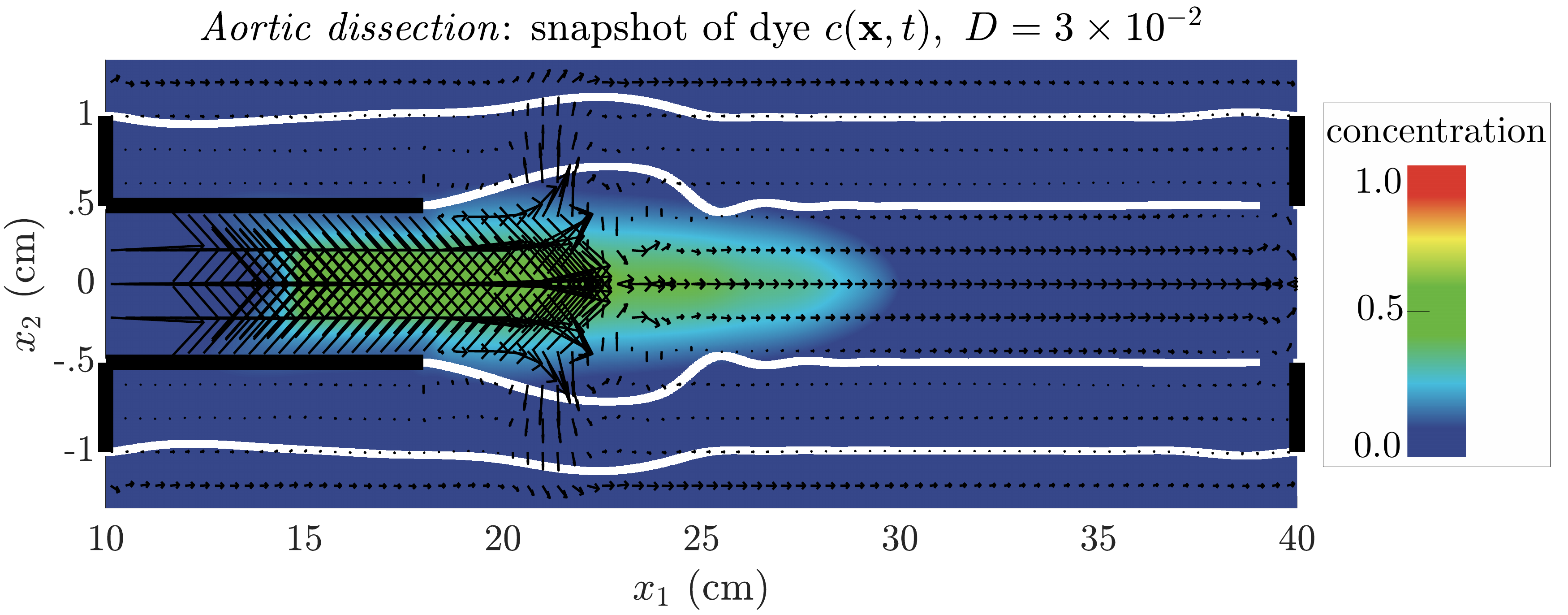}
\end{center}
\caption{\label{fig:fig_15}Numerical snapshots of the concentration after a dye injection for (top) zero physical diffusion and (bottom) non-zero physical diffusion ($D$ is in units of cm$^2$/s). Black lines indicate rigid walls, and white lines indicate compliant (elastic) walls.}
\end{figure}

\begin{figure}
  \begin{center}
\includegraphics[width=.72228\textwidth]{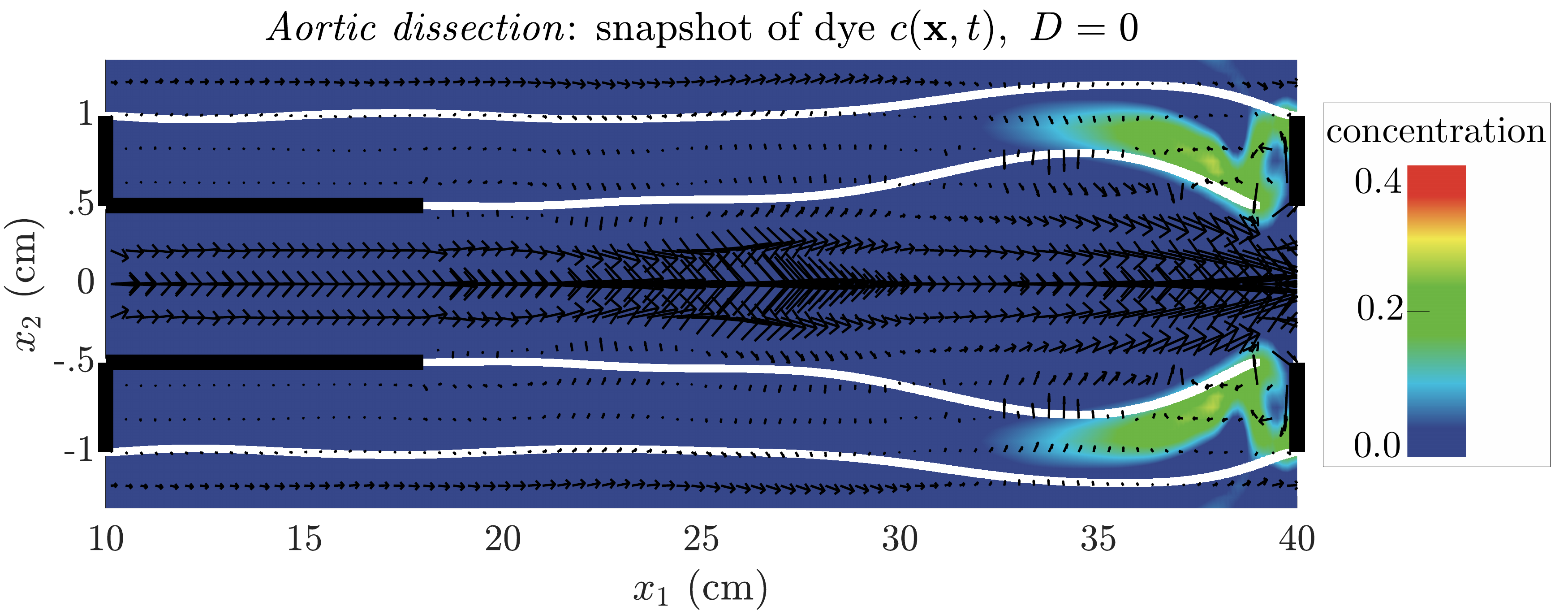}\\[1em]
\includegraphics[width=.72228\textwidth]{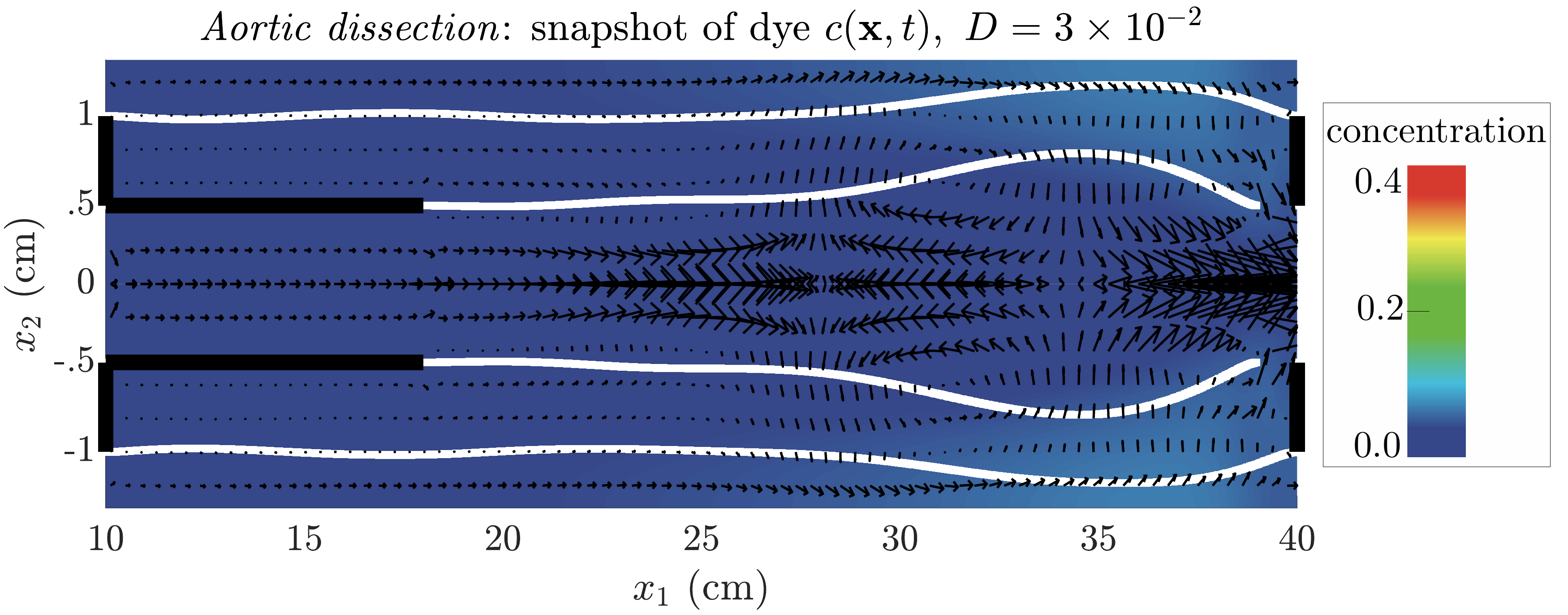}
\end{center}
\caption{\label{fig:fig_16}Numerical snapshots of the concentration several cardiac cycles after the dye injection ceases for (top) zero physical diffusion and (bottom) non-zero physical diffusion ($D$ is in units of cm$^2$/s). Black lines indicate rigid walls, and white lines indicate compliant (elastic) walls.}
\end{figure}

Figure~\ref{fig:fig_15} presents snapshots of an FC-based simulation of a dye injected across a small part (the central 0.5 cm) of the inlet of the dissected region (within the true lumen) for $5$ s (or 5 cycles) for both zero physical diffusion and non-zero physical diffusion ($D=D_{x_1}=D_{x_2} = 3\times10^{-2}$ cm$^2$/s). The latter demonstrates how a diffusive dye (or drug) fills up much of the true lumen of the aorta. Figure~\ref{fig:fig_16} presents corresponding snapshots several cardiac cycles after the dye injection ceases, where in both cases the dye passes through the free end of the septum, remaining trapped even after several additional cycles (a similar effect is seen in the PRT calculations of Figure~\ref{fig:fig_14}).

\subsection{\label{sec:fontan}Non-Newtonian flow through a Fontan circulation (3D)}

\begin{figure}
\includegraphics[width=.485\textwidth]{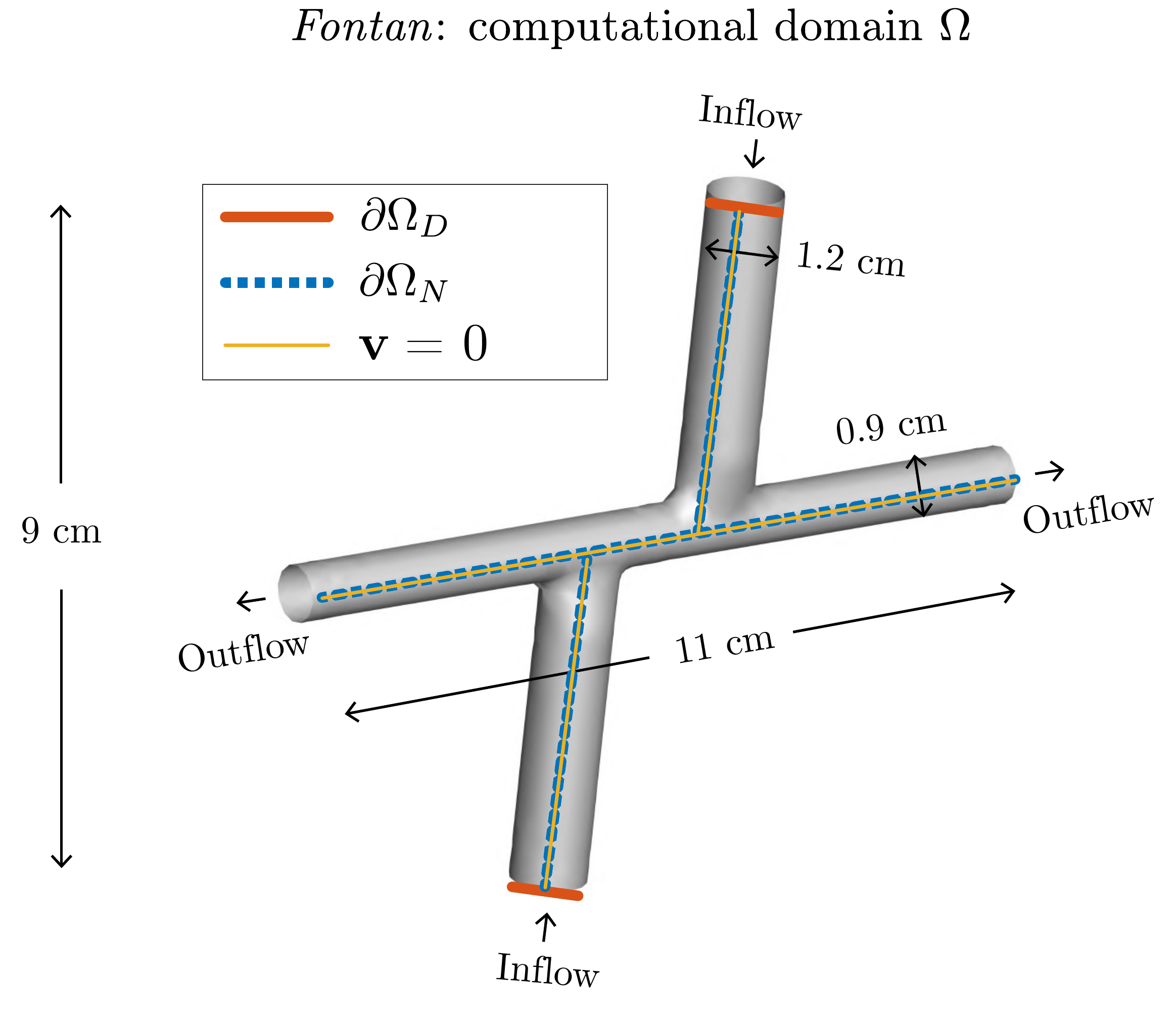}\hfill
\includegraphics[width=.485\textwidth]{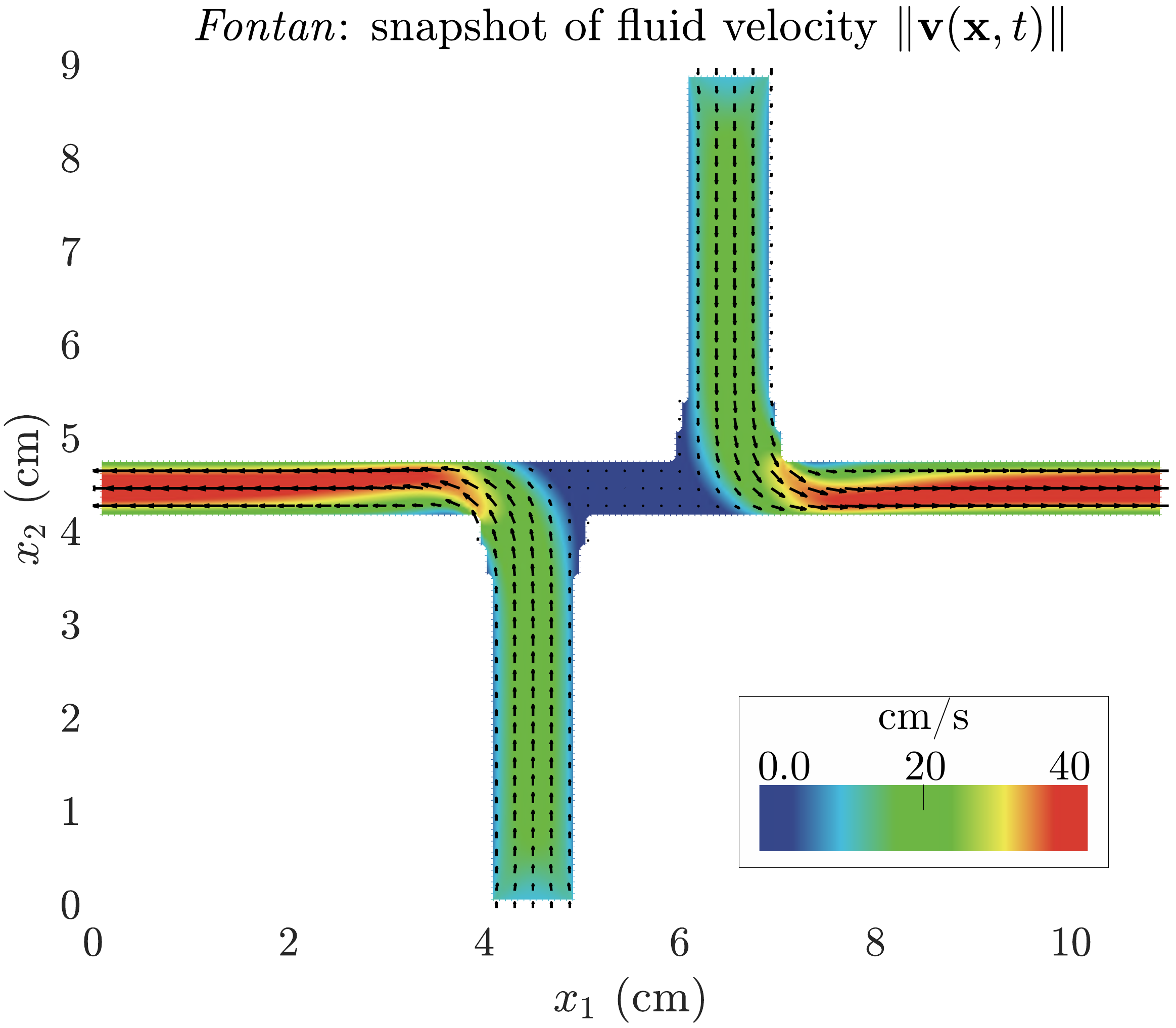}
\caption{\label{fig:fig_17}The computational configuration (left) for the Fontan circulation. (Right) A snapshot (depth mid-plane cross-section) of the corresponding steady non-Newtonian velocity field produced by the lattice Boltzmann method. }
\end{figure}

 Fontan patients don’t have the right side of the heart, and blood flow is pushed from superior vena cava (SVC) and the inferior vena cava (IVC) to the lungs in a passive way via a cross-shaped graft (known as a ``Fontan graft")~\cite{erbel1993effect}. The 3D computational domain $\Omega$ of a Fontan circulation is given in Figure~\ref{fig:fig_17} (left), where the top inflow is from the SVC and the bottom inflow is from the IVC. The two outlets flow towards the left and right lungs. The non-Newtonian fluid velocities are provided by the lattice Boltzmann solver described earlier with rigid walls and zero traction at the outlet (the shear rate viscosity curve of the shear-thinning Carreau–Yasuda model matching with Fontan patients is similar to~\cite{wei2020}). Blood fluid parameters correspond to $\mathrm{Re}=1515$, with a steady source corresponding to a cardiac output of 3 L/min. Further details of this configuration and its physiological and clinical relevancy are provided elsewhere~\cite{wei2020}. A snapshot (depth mid-plane cross-section) of the steady-state non-Newtonian flow velocity in the domain is presented in Figure~\ref{fig:fig_17} (right).

\begin{figure}
  \begin{center}
\includegraphics[width=.485\textwidth]{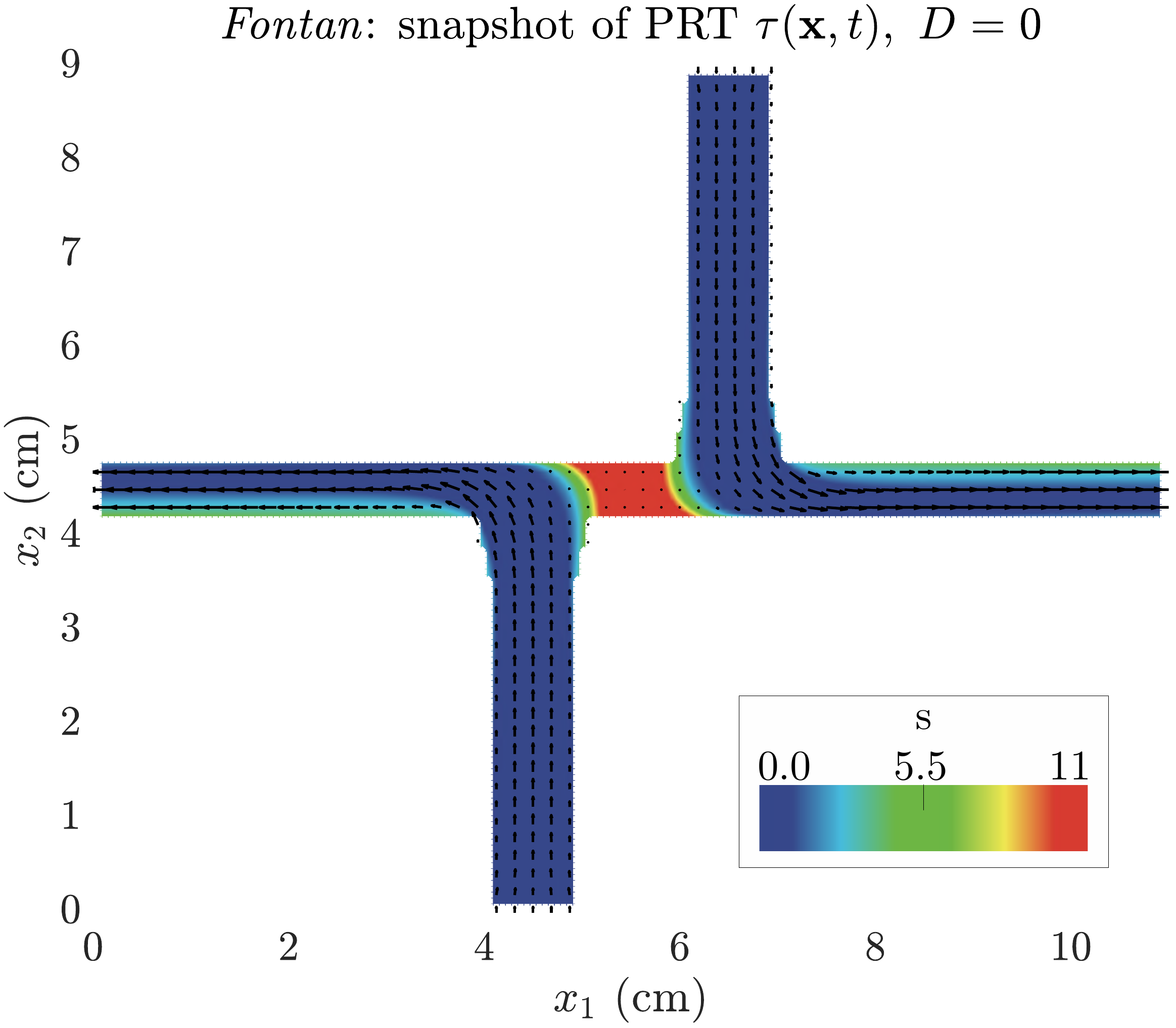}
\end{center}
\caption{\label{fig:fig_18}A numerical snapshot (depth mid-plane cross-section) of particle residence times for zero physical diffusion. Residence times is expectedly high in the central area between the two flows.}
\end{figure}

\begin{figure}
\includegraphics[width=.485\textwidth]{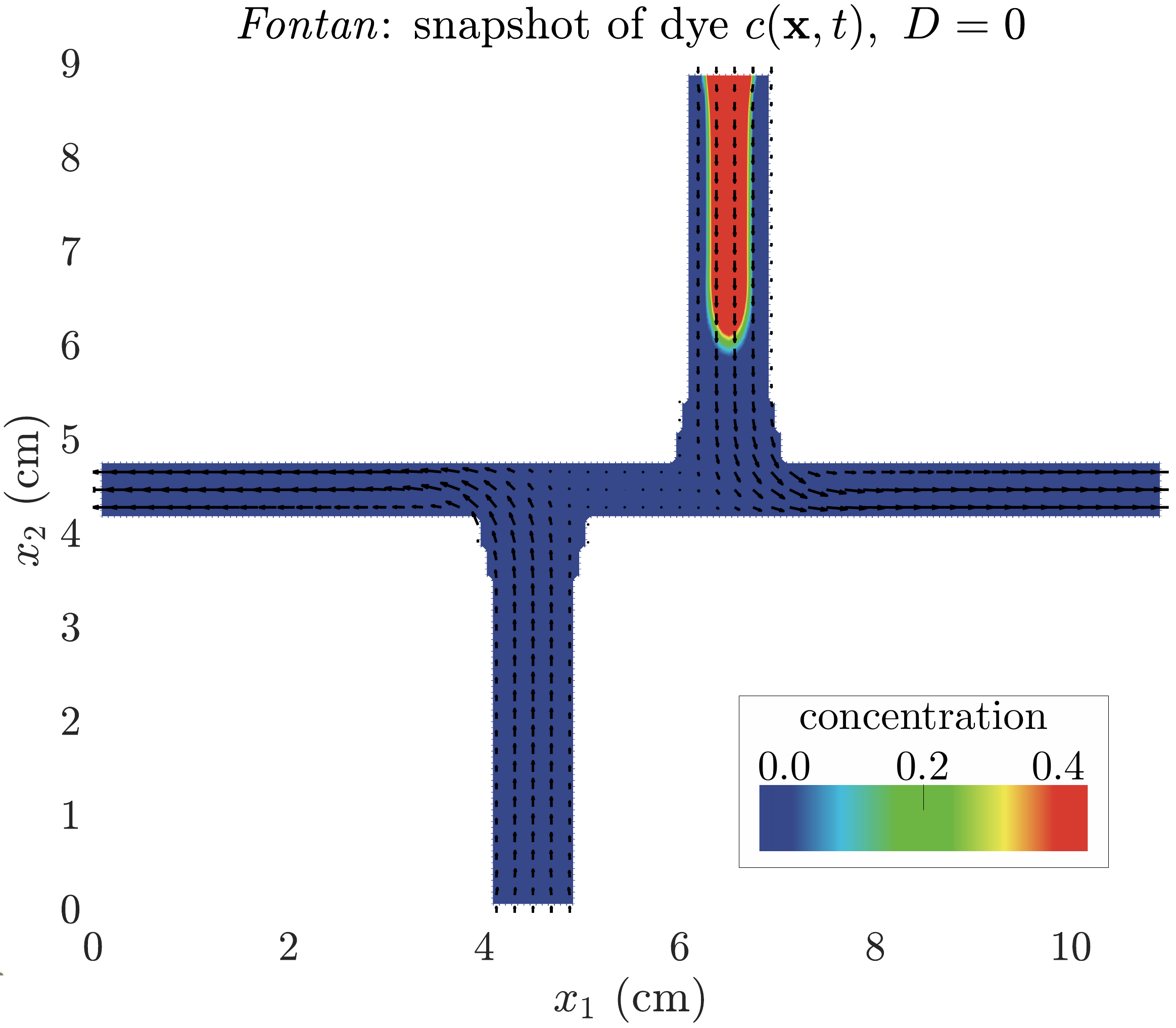}\hfill
\includegraphics[width=.485\textwidth]{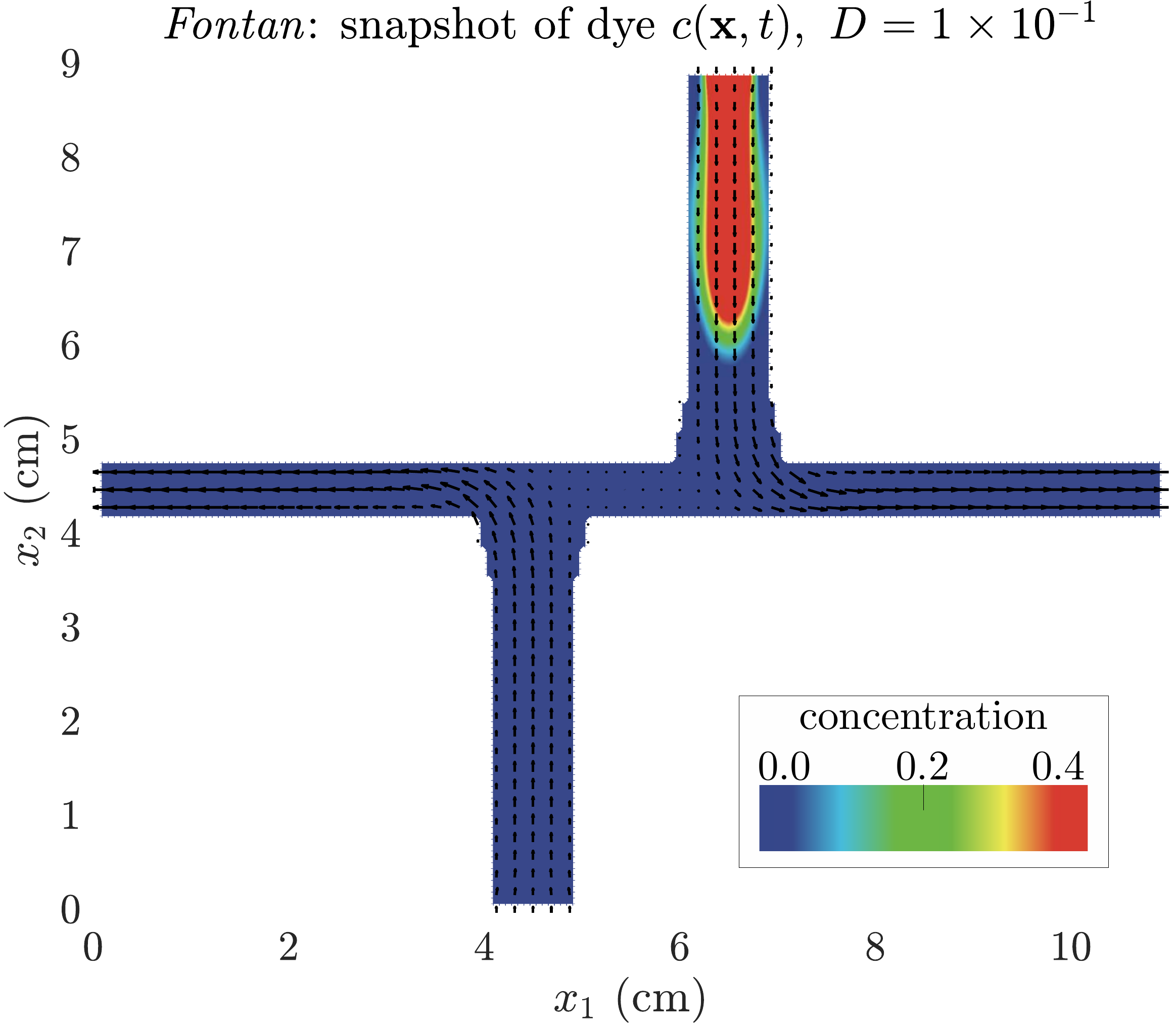}
\caption{\label{fig:fig_19}Numerical snapshots (depth mid-plane cross-section) of the concentration after a dye injection for (left) zero physical diffusion and (right) non-zero physical diffusion ($D$ is in units of cm$^2$/s). }
\end{figure}

\begin{figure}
\includegraphics[width=.485\textwidth]{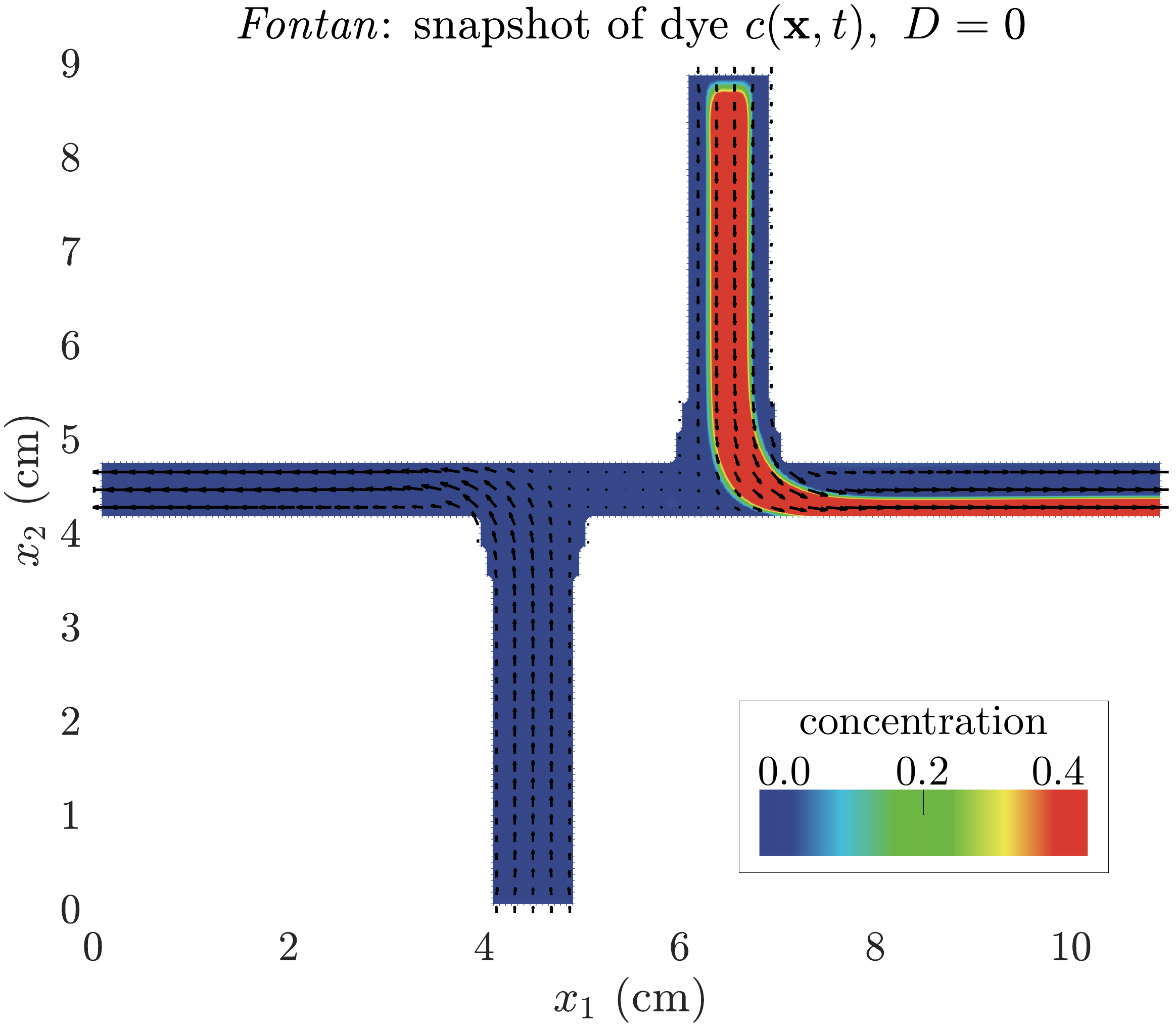}\hfill
\includegraphics[width=.485\textwidth]{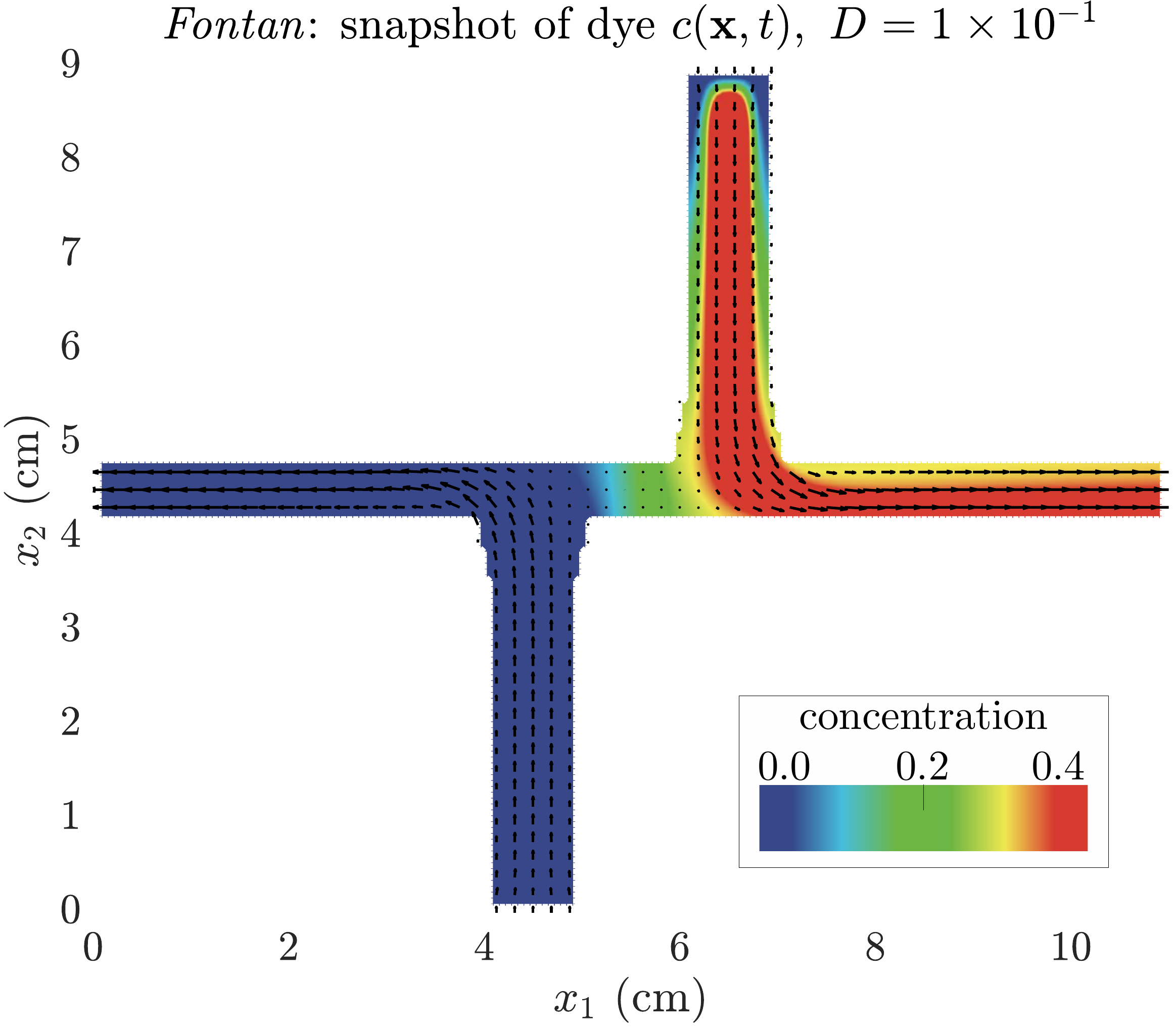}
\caption{\label{fig:fig_20}Numerical snapshots (depth mid-plane cross-section) of the concentration several cycles after a dye injection for (left) zero physical diffusion and (right) non-zero physical diffusion ($D$ is in units of cm$^2$/s). The physically-diffusive solution computed by FC demonstrates dye particles diffusing into the central region. }
\end{figure}

Using a spatial discretization of $176\times144\times16$ points (corresponding to $\Delta x_1 = \Delta x_2 = \Delta x_3 = 1/16$ cm) and a timestep of $\Delta t = 0.2$ ms for the convection-diffusion PDE, the FC-based particle residence time calculation for zero physical diffusion is presented in Figure~\ref{fig:fig_18}. The residence time is high at the area between the two outlets. Figure~\ref{fig:fig_19} presents snapshots during the initial cycles of an FC-based simulation of a dye injected across a subset of the top inlet for $20$ s (or 20 cycles) for both zero physical diffusion and non-zero physical diffusion ($D=D_{x_1}=D_{x_2} = 1\times10^{-1}$ cm$^2$/s). As expected, the physically-diffusive case begins to fill up the entire inlet of the Fontan graft. Figure~\ref{fig:fig_20} presents corresponding snapshots after several cycles, where in the physically-diffusive case, the dye particles diffuse into the area between the two outlets.

\section{\label{sec:conclusions}Conclusions}

This work {presents} the successful construction of a new pseudo-spectral (FC-based) methodology for conducting dye simulations and non-discrete PRT calculations. Studies analyzing errors and convergence (with analytical solutions of both forced and unforced systems) attest to high-order accuracy without the incurrence of numerical diffusion errors (which are well-known to exist for commonly-used (low-order) finite element or finite difference-based methods~\cite{kalinowska2007,sankar1998,mullen}). Successful applications of the new solver to physically-realistic and physiologically-relevant case studies demonstrate the efficacy and versatility of the proposed approach. {Although the FC method requires a regularity that precludes its direct application to shocks or compressible flow, previous work has demonstrated that FC-based solvers can be successfully coupled to shock-capturing schemes for building hybrid solvers that can treat a variety of conservation laws~\cite{sabh,shz}.} Although velocity data is provided in this work by numerical solutions (facilitated here by an FSI immersed boundary-lattice Boltzmann solver), the post-processing nature of the dye/PRT models can also apply to velocity data collected through suitable experimental techniques (e.g., particle image velocimetry) or clinical imaging modalities (e.g., cardiac 4D flow MRI or ultrasound image velocimetry). {However, the accuracy of such applications may be limited to the resolution and signal-to-noise ratio of acquired images and signals.} Analysis on the accuracy of the current method applied to such experimental data (e.g., compared with experimental dye tracing) is a subject of future work.


\bibliography{references.bib}

\end{document}